\documentclass[pra,aps,twocolumn,showpacs,floatfix,nofootinbib,superscriptaddress]{revtex4}

\usepackage{amsmath,amssymb}
\usepackage{dcolumn}
\usepackage{bm}
\usepackage{graphicx}
\usepackage{subfigure}

\graphicspath{{figures/pdf/},{figures/eps/}}
\DeclareGraphicsExtensions{.pdf}

\def\vec#1{\mathbf{#1}}

\def\ket#1{| #1 \rangle}

\def\ave#1{\langle #1 \rangle}
\def\norm#1{|| #1 ||}
\def\sx{\sigma_x}

\def\sz{\sigma_z}

\def\D{\mathcal{D}}
\def\E{\mathcal{E}}

\def\diag{\mbox{\rm diag}}

\def\est{{\rm est}}
\def\Var{\mathop{\rm Var}}

\begin{document}
\title{The ubiquitous problem of learning system parameters for
  dissipative two-level quantum systems: Fourier analysis
  versus Bayesian estimation}

\author{Sophie G.\ Schirmer}
\affiliation{College of Science (Physics), Swansea University,
  Singleton Park, Swansea, SA2 8PP, United Kingdom}
\author{Frank C.\ Langbein}
\affiliation{College of Physical Sciences \& Engineering
 (Computer Science \& Informatics), Cardiff University,
  5 The Parade, Cardiff, CF24 3AA, United Kingdom}
\date{\today}

\begin{abstract}
We compare the accuracy, precision and reliability of different
methods for estimating key system parameters for two-level systems
subject to Hamiltonian evolution and decoherence. It is demonstrated
that the use of Bayesian modelling and maximum likelihood estimation
is superior to common techniques based on Fourier analysis.  Even for
simple two-parameter estimation problems, the Bayesian approach yields
higher accuracy and precision for the parameter estimates obtained. It
requires less data, is more flexible in dealing with different model
systems, can deal better with uncertainty in initial conditions and
measurements, and enables adaptive refinement of the estimates. The
comparison results shows that this holds for measurements of large
ensembles of spins and atoms limited by Gaussian noise as well as
projection noise limited data from repeated single-shot measurements
of a single quantum device.
\end{abstract}


\maketitle

\section{Introduction}

Quantum systems play an important role in atomic and molecular
physics, chemistry, material science and many important current
technologies such as nuclear magnetic resonance imaging~\cite{MRI} and
spectroscopy~\cite{MRS}, promising nascent quantum technologies such
as spintronic devices~\cite{Spintronics}, and potential future
technologies such as quantum information processing~\cite{QIP}.  Novel
applications require increasingly sophisticated control, and accurate
and precise models to facilitate controlled manipulation of their
dynamics.

Although theoretical device modelling remains important, system
identification and data-driven models are becoming increasingly
important in many areas of science and technology to accurately
describe individual systems~\cite{Data-driven}. System identification
comprises a range of problems including model identification, model
discrimination and model verification.  Once a model has been
selected, the task often reduces to identifying parameters in the
model from experimental data.  In the quantum domain this is often
data from one of the many types of spectroscopy, from magnetic
resonance to laser to electron transmission spectroscopy, depending on
the physical system. More recently single shot measurements of quantum
systems have also become important for quantum devices relying on
individual quantum states.

Fourier analysis of the spectra is frequently used to identify model
parameters such as chemical shifts and relaxation rates by examination
of the positions and shape of peaks in a free-induction-decay (FID)
spectrum~\cite{NMR}.  Fourier analysis of Rabi oscillation spectra has
also been used to identify Hamiltonians~\cite{Schirmer2004, Cole2005},
as well as decoherence and relaxation parameters for two-level
systems~\cite{Cole2006}, and concurrence spectroscopy~\cite{Cole2006b}
has been applied to determine information about coupling between
qubits.  For more complex systems, Bayesian techniques and maximum
likelihood estimation~\cite{Bretthorst} have proved to be extremely
valuable to construct data-driven models to identify Hamiltonian
parameters~\cite{Schirmer2009} and decoherence parameters for
multi-level systems~\cite{Oi2010}.  Bayesian techniques have also been
applied for adaptive Hamiltonian learning using sequential Monte-Carlo
techniques~\cite{Granade2012}.

In this work we revisit simpler systems: two-level systems subject to
decoherence, one of the simplest but arguably most important models in
quantum physics.  The model is ubiquitous in magnetic resonance
imaging, where the magnetization signal from protons
(spin-$\tfrac{1}{2}$ particles) precessing and dephasing in a magnetic
field is the basis for non-invasive, in-vivo imaging.  In quantum
information it describes qubits as the fundamental building blocks
subject to decoherence.  Therefore, characterization of two-level
systems is extremely important.  We compare two frequently used
estimation strategies based on Fourier analysis and a Bayesian
approach combined with maximum likelihood estimation, for the
ubiquitous parameter estimation problem of a two-level system subject
to decoherence.  We consider accuracy, precision and efficiency for
different systems and noise models, including Gaussian noise,
typically encountered for large ensemble measurements, and projection
noise, typically present in data from repeated single-system
measurements.

\section{System and Experimental Assumptions}

In this section we introduce our dynamic model of the physical system
and our assumptions about initialisation and measurement of the
system.  We focus in particular on the different options for the
measurements depending on the nature of the physical system and hence
the measurements from which we wish to estimate the parameters.

\subsection{Dynamic system model}

The state of a quantum system is generally described by a density
operator $\rho$, which, for a system subject to a Markovian
environment, evolves according to a Lindblad-type master equation
\begin{equation}\label{eq:model}
  \begin{split}
  \dot{\rho}(t) &= [H_0,\rho(t)] + \D[V]\rho, \\
  \D[V]         & = V \rho V^\dag - \tfrac{1}{2} (V^\dag V \rho + \rho V^\dag V),
   \end{split}
\end{equation}
where $H$ represents the Hamiltonian and $V$ the dephasing operator.
If the dephasing occurs in the same basis as the Hamiltonian evolution
then we can choose a basis in which both $H_0$ and $V$ are diagonal.
For a two-level system we can thus write $H=\omega\sz$ and
$V=\tilde{\gamma}\sz$, where $\tilde{\gamma} \ge 0$, leaving us
essentially with two core system parameters to identify, $\omega$ and
$\tilde{\gamma}$, or often $\gamma=2\tilde{\gamma}^2$.

\subsection{Initialization and Readout}

A basic experiment involves initalizing the system in some state
$\ket{\psi_I}$ and measuring the decay signal, a so-called
free-induction decay experiment.  The measured signal depends on the
system parameters as well as the initial state and the measurement.
Taking the measurement operator to be of the form
\begin{equation}
    M = \begin{pmatrix} \cos\theta_M &  \sin\theta_M  \\
                        \sin\theta_M & -\cos\theta_M
        \end{pmatrix},
\end{equation}
and taking the initial state to be
\begin{equation}
   \ket{\psi_I} =\cos(\theta_I)\ket{0}+\sin(\theta_I)\ket{1},
\end{equation}
the measurement signal is of the form
\begin{equation}
\label{eq:meas1}
  p(t) = e^{-\gamma{t}}\cos(\omega t)\sin(\theta_I)\sin(\theta_M) +\cos(\theta_I)\cos(\theta_M).
\end{equation}
Assuming the system is initially in the ground state $\ket{0}$, e.g.,
corresponding to spins being aligned with an external magnetic field,
the initialization procedure corresponds to applying a short pulse to
put the system into a superposition of the ground and excitation
state.  Notice if the system is not well characterized then it is
likely to be infeasible to prepare the system in a well-defined
superposition state with a known angle $\theta_I$.  Rather, $\theta_I$
becomes an additional parameter to be estimated.

The operator $M$ corresponds to measuring the system with regard to an
axis tilted by an angle $\theta_M$ from the system axis in the $(x,z)$
plane, which can describe many different experimental situations.  In
an FID experiment in NMR, for example, an $x$-magnetization
measurement corresponds to setting $\theta_M=\frac{\pi}{2}$.  In a
Rabi spectroscopy experiment of a quantum dot, where the population of
the ground and/or excited state is measured, e.g., via a fluorescence
measurement, we would typically set $\theta_M=0$.  In some situations,
such as the examples mentioned, the Hamiltonian and measurement bases
may be well-known.  In other situations, however, such as in a double
quantum dot system with charge state read-out via a single electron
transistor perhaps, $\theta_M$ may a priori at most be approximately
known.  In this case $\theta_I$ becomes an additional parameter to be
estimated.  In this work we employ a formalism that does not require
either the initial state or measurement to be known a priori.

\subsection{Continuous vs discrete-time and adaptive measurements}

In an FID experiment we could in principle measure the decay signal
continuously.  However, modern receivers typically return a digitized
signal, i.e., a vector of time samples, usually the signal values
integrated over short time intervals $\Delta t$. For this type of
readout, the number $N$ of time samples and their spacing $\Delta t$
are usually fixed, or at least selected prior to the start of the
experiment.  In this set-up there is usually little opportunity for
adaptive refinement short of simply repeating the entire experiment
with shorter $\Delta t$ or larger $N$.

In other situations, such as Rabi spectroscopy~\cite{RabiSpect}, each
measurement corresponds to a separate experiment.  For example, we
prepare the system in a certain initial state, let it evolve under
some Hamiltonian (with parameters to be estimated) for some time $t$
before performing a measurement to determine the state of the system.
In this case we are more flexible and can in principle choose the
measurement times adaptively, trying to optimize the times to maximize
the amount of information obtained in each measurement.

Here we mainly consider the case of a regularly sampled measurement
signal but we also briefly consider how the estimation can be improved
in the latter case by adaptive sampling with particular focus on the
comparison between the different estimation strategies.

\subsection{Ensemble vs single-system measurements}

In many settings from NMR and MRI to electron spin resonance (ESR) to
atomic ensembles in atom traps, large ensembles of spins or atoms are
studied resulting in ensemble average measurements.  In this setting,
the backaction from the measurement is negligible and the system can
be measured continuously to obtain a measurement signal $s(t)$.  The
noise in the signal is well approximated by Gaussian noise, which can
be simulated by adding a zero-mean Gaussian noise signal $g(t)$ to the
ideal signal $p(t)$, i.e., the measured signal $d(t)=p(t)+g(t)$. By
the Law of Large Numbers and Iterated Logarithm Law~\cite{Feller1968}
this gives a Gaussian distribution for $d(t)$ with mean $p(t)$ and
variance $\sigma^2\sim \tfrac{\log\log N_e}{2 N_e}$ for
$N_e\to\infty$.  This is a good error model for simulating physical
systems and estimating the noise in actual measurement data when the
ensemble size $N_e$ is large.

More recently single quantum systems, such as trapped
ions~\cite{Ions}, trapped atoms~\cite{Atoms}, single electron
spins~\cite{SingleSpin}, and charge states in Josephson
junctions~\cite{Josephson}, have become an important topic for
research because of their potential relevance to quantum technolgoies.
Given a single copy of a two-level system, measurement of any
observable yields only a single bit of information indicating a 0 or 1
result. To determine the expectation value of an observable the
experiment has to be repeated many times and the results averaged.
Furthermore, due to the backaction of the measurement on the system,
we can generally only perform a single projective measurement.  To
obtain data about the observable at different times the system has to
be re-initialized and the experiment repeated for each measurement.
In this context the ensemble size $N_e$ is the number of times each
experiment on a single copy of the system is repeated.  As repetitions
are time- and resource-intensive, it is desirable to keep $N_e$
small. However, this means the precision of the expectation values of
observables becomes limited by projection noise, following a Poisson
distribution.  To simulate experiments of this type we compute the
probability $\hat{p}_1$ of measurement outcome $1$ for the simulated
system, generate $N_e$ random numbers $r_n$ between $0$ and $1$, drawn
from a uniform distribution, and set $p_1=N_1/N_e$, where $N_1$ is the
number of $r_n \le \hat{p}_1$.

\section{Parameter Estimation Strategies}

This section introduces the three parameter estimation strategies
based on Fourier and Bayesian analysis we wish to compare.

\subsection{Fourier-spectrum based estimation}

A common technique to find frequency components in a noisy time-domain
signal is spectral analysis.  Consider a measurement signal of the form
\begin{equation}
   \label{eq:meas}
   p(t) = a + b e^{-\gamma t}\cos(\omega_0 t), \qquad t\ge 0,
\end{equation}
which corresponds directly to measurement (\ref{eq:meas1}) if we set
$a=\cos\theta_I\cos\theta_M$ and $b=\sin\theta_I\sin\theta_M$.
Subtracting the mean of the signal $\ave{p(t)}=a$ and rescaling gives
$f(t)=(p(t)-a)/b$. To account for the fact that $f(t)$ is defined only
for $t\ge 0$ we multiply $f(t)$ by the Heaviside function
\begin{equation*}
  u(t) = \begin{cases} 0 & \mbox{if } t <0 \\
                       1 & \mbox{if } t \ge 0.
         \end{cases}
\end{equation*}
The Fourier transform of $u(t)f(t) = u(t)e^{-\gamma t}\cos(\omega_0t)$ is
\begin{equation*}
   F(\omega) = \frac{\gamma + i\omega}{(\gamma+i\omega)^2+\omega_0^2}
\end{equation*}
and the power spectrum is $P(\omega) = |F(\omega)|^2$. Differentiating
with respect to $\omega$ and setting the numerator to $0$ shows that
$|F(\omega)|^2$ has extrema for $\omega=0$ and
$(\gamma^2+\omega^2)^2-\omega_0^2(4\gamma^2+\omega_0^2)=0$.  The real
roots $\omega_*$ of this equation satisfy
\begin{equation}
   \E_1(\omega_0,\gamma) = \omega_*^2 + \gamma^2 -\omega_0\sqrt{4\gamma^2+\omega_0^2}=0
\end{equation}
and the corresponding maximum of the power spectrum
\begin{equation*}
 P_* = P(\omega_*)
     = \frac{\omega_0^2+\omega_0\sqrt{4\gamma^2+\omega_0^2}}{8\gamma^2\omega_0^2}
     = \frac{\omega_0^2+\omega_*^2+\gamma^2}{8\gamma^2\omega_0^2}.
\end{equation*}
Defining the error term
\begin{equation}
  \E_2(\omega_0,\gamma) = 8\gamma^2\omega_0^2 P_*-\omega_0^2 + \gamma^2 + \omega_*^2,
\end{equation}
we can estimate the frequency $\omega_0$ and dephasing rate $\gamma$
from the peak height $P_*$ and position $\omega_*$ via
\begin{align}
  &\mbox{\textbf{Strategy 1: }} \nonumber\\
  &\{\omega_0,\gamma\}  = \arg\min_{\omega_0',\gamma'} & \{ |\E_1(\omega_0',\gamma')|+|\E_2(\omega_0',\gamma')|\}.
\end{align}
Determining the maximum $P_*$ and its location $\omega_*$ from
$|F(w)|^2$, we may choose $\omega_0'=\omega_*$ and $\gamma' =
\sqrt{2\omega_*/(8\omega_*^2 P_* -1)}$ as starting point for a local
minimization routine provided $\gamma\ll \omega_0$ as is usually the
case.

Instead of estimating the height of the peak, estimates for $\omega_0$
and $\gamma$ can also be obtained using the width of the peak.  Let
$\omega_{1,2}$ be the (positive) frequencies for which $|F(\omega)|$
assumes half its maximum. One way to estimate $\omega_{1,2}$ is to
take the minimum and maximum of $\{ \omega : |F(\omega) \geq \max(F)
\}$, assuming that sufficient measurements have been made such that
$F$ is symmetric and peaked, i.e., it has low skewness and high
kurtosis.

The full-width-half-maximum $2d$ of $|F(\omega)|$ is
$|\omega_2-\omega_1|$ and we can derive the following expression:
\begin{equation*}
\begin{split}
  d &= \left[\sqrt{\omega_0^2-\gamma^2+2\sqrt{3}\omega_0\gamma}-\sqrt{\omega_0^2-\gamma^2} \right]\\
    &= \left[\sqrt{\omega_*^2+2\sqrt{3} \gamma\sqrt{\omega_*^2+\gamma^2}}-\omega_* \right].
\end{split}
\end{equation*}
Hence, given the location $\omega_*$ and half-width $d$ of the peak
solving for $\gamma$ gives the alternative
\begin{align}
  & \mbox{\textbf{Strategy 2: }} \nonumber\\
  & \gamma = \frac{1}{6} \sqrt{6 g(\omega_*,d)-18\omega_*^2}, &&
  \omega_0=\sqrt{\omega_*^2+\gamma^2},
\end{align}
where $g(\omega_*,d)=\sqrt{9 \omega_*^4 + 12 d^2 \omega_*^2 + 12 d^3
\omega_* + 3 d^4}$.

Strategy 2 based on peak-positions and linewidths is probably the most
common approach for estimating frequencies and $R_2$-relaxation rates
from FID signals in NMR and in many other contexts.  The expressions
for $|P(\omega)|^2$, the peak heights and linewidth are more
complicated than those for quadrature measurements as we only have a
real cosine signal but the approach is fundamentally the same.

\subsection{Bayesian and Maximum Likelihood Approach}

Given discrete time-sampled data represented by a row vector $\vec{d}$
of length $N_t$ containing the measurement results obtained at times
$t_n$ for $n=1, \ldots, N_t$, let $\vec{p}$ be the vector of the
corresponding measurement outcomes predicted by the model. $\vec{p}$
depends on the model parameters, here $\omega_0$ and $\gamma$.
Assuming Gaussian noise with variance $\sigma^2$ we define the joint
likelihood~\cite{Bretthorst}
\begin{equation}
  \label{eq:likelihood}
  P(\vec{p},\vec{d},\sigma) =
  \frac{1}{(\sqrt{2\pi}\sigma)^{N_t}}\exp\left[-\frac{\norm{\vec{p}-\vec{d}}_2^2}{2\sigma^2}\right].
\end{equation}
If the noise level $\sigma$ of the data is not known a priori, we can
eliminate this parameter following the standard Bayesian approach by
integrating over $\sigma$ from $0$ to $\infty$, using the Jeffrey's
prior $\sigma^{-1}$. This gives
\begin{equation}
 \label{eq:likelihood2}
 P(\vec{p},\vec{d}) = \frac{\Gamma(\tfrac{N_t}{2}-1)}{4\pi^{N_t/2}} \norm{\vec{p}-\vec{d}}_2^{2 - N_t}
\end{equation}
where $\Gamma$ is the Gamma function.  It is usually more convenient
and numerically robust to work with the (negative) logarithm of the
likelihood function, the so-called log-likelihood.  When the noise
level $\sigma$ is known the log-likelihood reduces to
\begin{equation}
 \label{eq:loglikelihood1}
  L(\vec{p},\vec{d},\sigma)
  = -\log P(\vec{p},\vec{d},\sigma)
  = \tfrac{1}{2\sigma^2} \norm{\vec{p}-\vec{d}}_2^2 + \mbox{const},
\end{equation}
where the constant is usually omitted; when $\sigma$ is not known a
priori we obtain instead
\begin{equation}
 \label{eq:loglikelihood2}
   L(\vec{p},\vec{d})
   = -\log P(\vec{p},\vec{d})
   = \frac{1-N_t}{2} \log \norm{\vec{p}-\vec{d}}_2^2 + \mbox{const}.
\end{equation}
The idea of maximum likelihood estimation is to find the model
parameters that maximize this (log-)likelihood function.  To simplify
this task, we follow a similar approach as in previous work
\cite{Bretthorst, Schirmer2009, Oi2010} and express the signals as
linear combinations of a small number $m_b$ of basis functions
$g_m(t)$ determined by the functional form of the signals.  In our
case the measurement signal $p(t)$ can be written as a linear
combination of $m_b=2$ basis functions
\begin{equation}\label{eq:im}
   p(t) = \alpha_1 g_1(t) + \alpha_2 g_2(t).
\end{equation}
with $g_1(t)=1$ and $g_2(t)=e^{-\gamma t}\cos(\omega_0 t)$.  As the
basis functions are not orthogonal, we define an orthogonal projection
of the data onto the basis functions sampled at times $t_n$ as
follows.  Let $G$ be a matrix whose rows are the basis functions
$g_m(t)$ evaluated at times $t_n$, $G_{mn}=g_m(t_n)$, and $E
\,\diag(\alpha_m)\, E^\dag$ be the eigendecomposition of the
positive-definite matrix $GG^\dag$.  Then
$H=\diag(\alpha_m^{-1/2})E^\dag G$ is a matrix satisfying $H^\dag
H=GG^\dag$, whose rows form an orthonormal set, $H H^\dag = I$, and we
define the orthogonal projection of the data vectors onto the basis
function by $\vec{h} = H\vec{d}^\dag$.

Projecting the data onto a linear combination of basis functions
introduced $m_b$ nuisance parameters $\alpha_m$.  Using a standard
Bayesian approach we can eliminate them by integration using a uniform
prior, and following further simplifications~\cite{Bretthorst}, it can
be shown that the log-likelihood~(\ref{eq:likelihood2}) becomes
\begin{equation}
 \label{eq:loglikelihood3}
   L(\omega_0,\gamma|\vec{d}) =
   \frac{m_b-N_t}{2} \log \left[1-\frac{m_b \ave{\vec{h}^2}}{N_t \ave{\vec{d}^2}} \right]
\end{equation}
where $\ave{\vec{d}^2} = \tfrac{1}{N_t} \sum_{n=0}^{N_t-1} d_n^2$ and
$\ave{\vec{h}^2} = \tfrac{1}{m_b} \sum_{m=0}^{m_b-1} h_m^2$ and we
have dropped the constant offset.  This log-likelihood function can be
evaluated efficiently, and we can use standard optimization algorithms
to find its maximum, motivating
\begin{align}
  \mbox{\textbf{Strategy 3: }} &&
  \{\omega_0, \gamma\} = \arg\max_{\omega_0',\gamma'} L(\omega_0',\gamma'|\vec{d}). &&
\end{align}
Note that in general, finding the global maximum of the log-likelihood
function is non-trivial as it is non-convex, tends to become sharply
peaked, especially for large data sets, and may have many local
extrema, necessitating global search techniques.  However, for our
two-parameter case, finding the global optimum over reasonable ranges
for $\omega$ and $\gamma$ proved straightforward using either standard
quasi-Newton or even Nelder-Mead Simplex optimization. For more
complex functions a density estimator such as particle filters
(sequential Monte Carlo methods) or kernel density estimators may be
used, which also enable effective determination of the maximum.

\begin{table*}
\scalebox{1}{
\begin{tabular}{|l|*{10}{c|}}
\hline
$\omega$ & 1.0000 & 0.9000 & 0.5003 & 0.7304 & 1.2161 & 1.6211 & 0.2218 & 1.5195 & 0.7551 & 0.8029 \\\hline
$\gamma$ & 0.1000 & 0.1000 & 0.1243 & 0.1875 & 0.2031 & 0.0993 & 0.1234 & 0.0751 & 0.0533 & 0.1921 \\\hline
\end{tabular}}
\caption{Model parameters for 10 models compared below (in units of $\bar{\omega}$).}
\label{table:models}
\end{table*}

\section{Evaluation and Comparison of Estimation Strategies}

We now compare the three strategies introduced in the previous section
for ensemble and single-shot measurements and also discuss the
uncertainty in the estimated parameters and show how Strategy~3
enables the estimation of additional initialisation and measurement
parameters. For this we use $10$ systems with different values for
$\omega$ and $\gamma$, given in Table~\ref{table:models}, and collect
measurement data from simulations with the relevant noise models. For
each system the signal was sampled uniformly at $N_t=100$ time points
$t_k \in [0,30]$. We assume that we have some order of magnitude
estimate of the system frequency $\bar{\omega}$ based on the physical
properties of the system, giving us a range for the values of
$\omega$.  Without loss of generality we can express both $\omega$ and
$\gamma$ in units of $\bar{\omega}$.  Accordingly all times quoted in
the following will be in units of $\bar{\omega}^{-1}$.  In our
simulations we choose $\omega \in [0.2,2]$ and $\gamma \in [0.05,0.4]$
in units of $\bar{\omega}$.

To calculate an average relative error for the parameter estimates,
$N_s=1000$ runs were performed for each system and noise level and the
error computed as
\begin{subequations}
  \begin{align}
  e(\omega) &= \frac{1}{N_s}\sum_{n=1}^{N_s} \omega^{-1}|\omega_{\est}^{(n)}-\omega| \\
  e(\gamma) &= \frac{1}{N_s}\sum_{n=1}^{N_s} \gamma^{-1}|\gamma_{\est}^{(n)}-\gamma|
\end{align}
\label{eq:error}
\end{subequations}
where $\omega$ and $\gamma$ are the actual parameters of the simulated
system and $\omega_{\est}^{(n)}$ and $\gamma_{\est}^{(n)}$ are the
estimated values for the $n$th run.

\begin{figure*}
\includegraphics[width=0.49\textwidth]{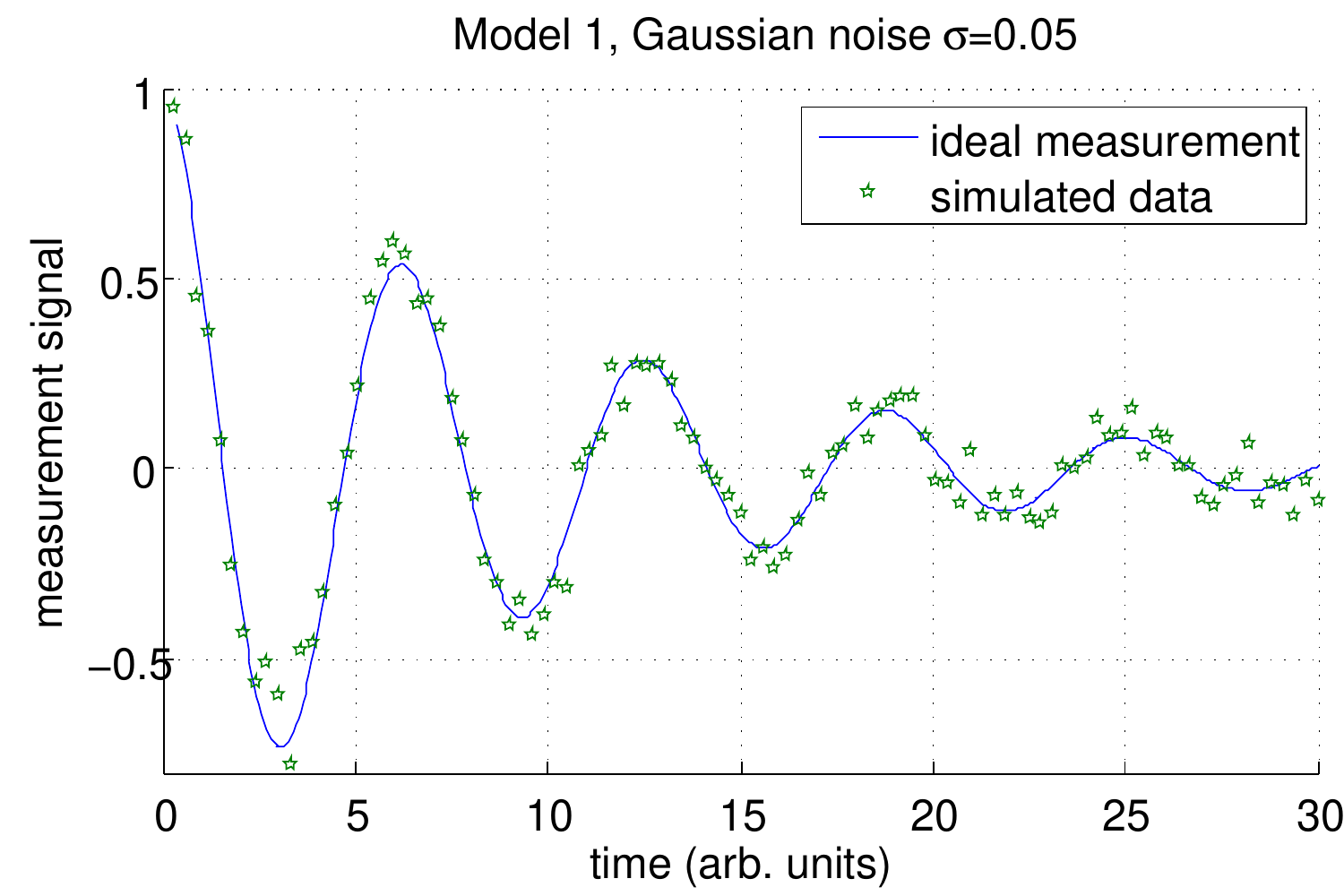} \hfill
\includegraphics[width=0.49\textwidth]{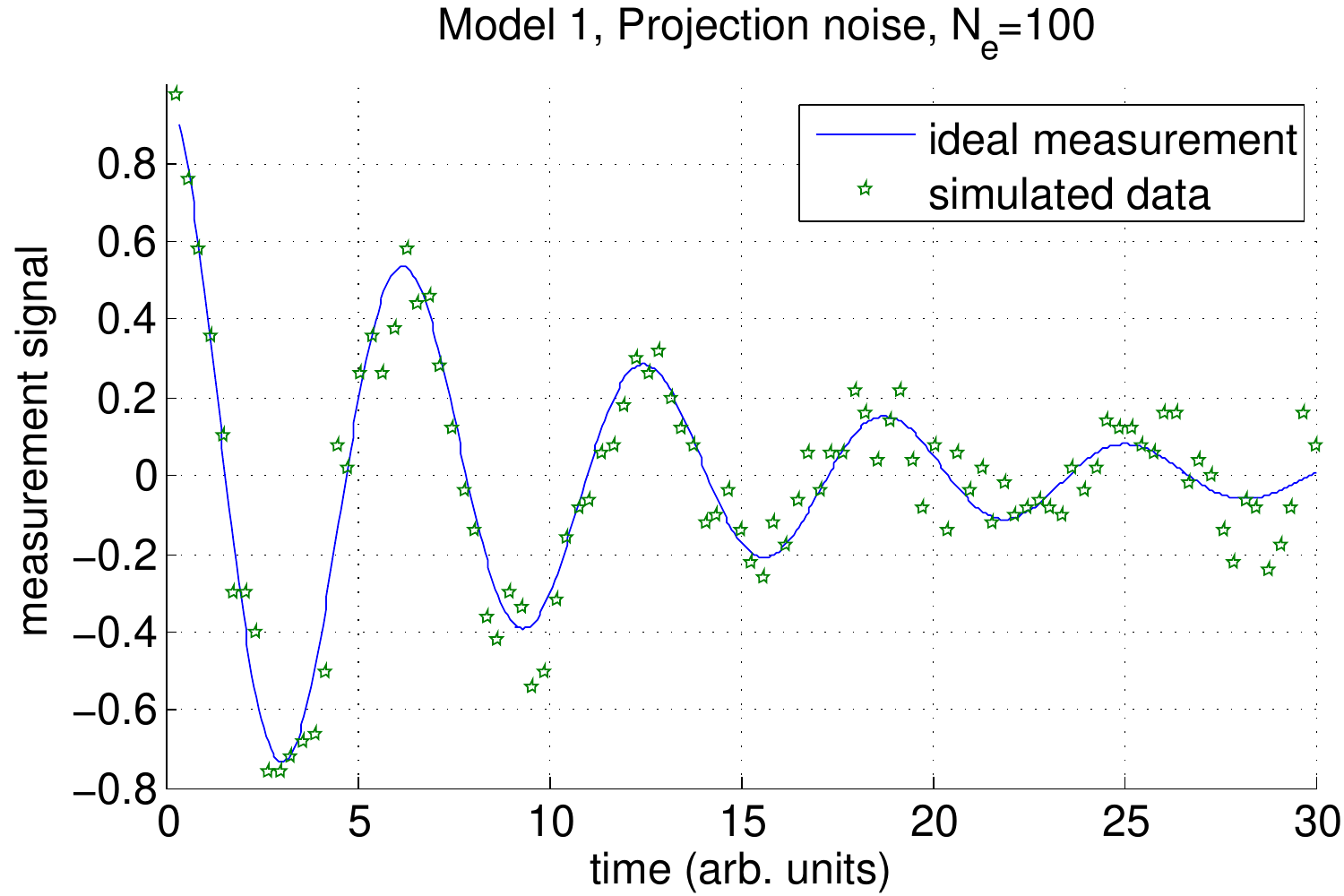}
\caption{Example of ideal measurement signal and data from simulated
  experiments with Gaussian noise ($\sigma=0.05$, left) and projection noise
  (each data point is the average of $N_e=100$ binary-outcome single
  shot measurements, right).}
\label{fig:signal}
\end{figure*}

\begin{figure*}
\includegraphics[width=0.49\textwidth]{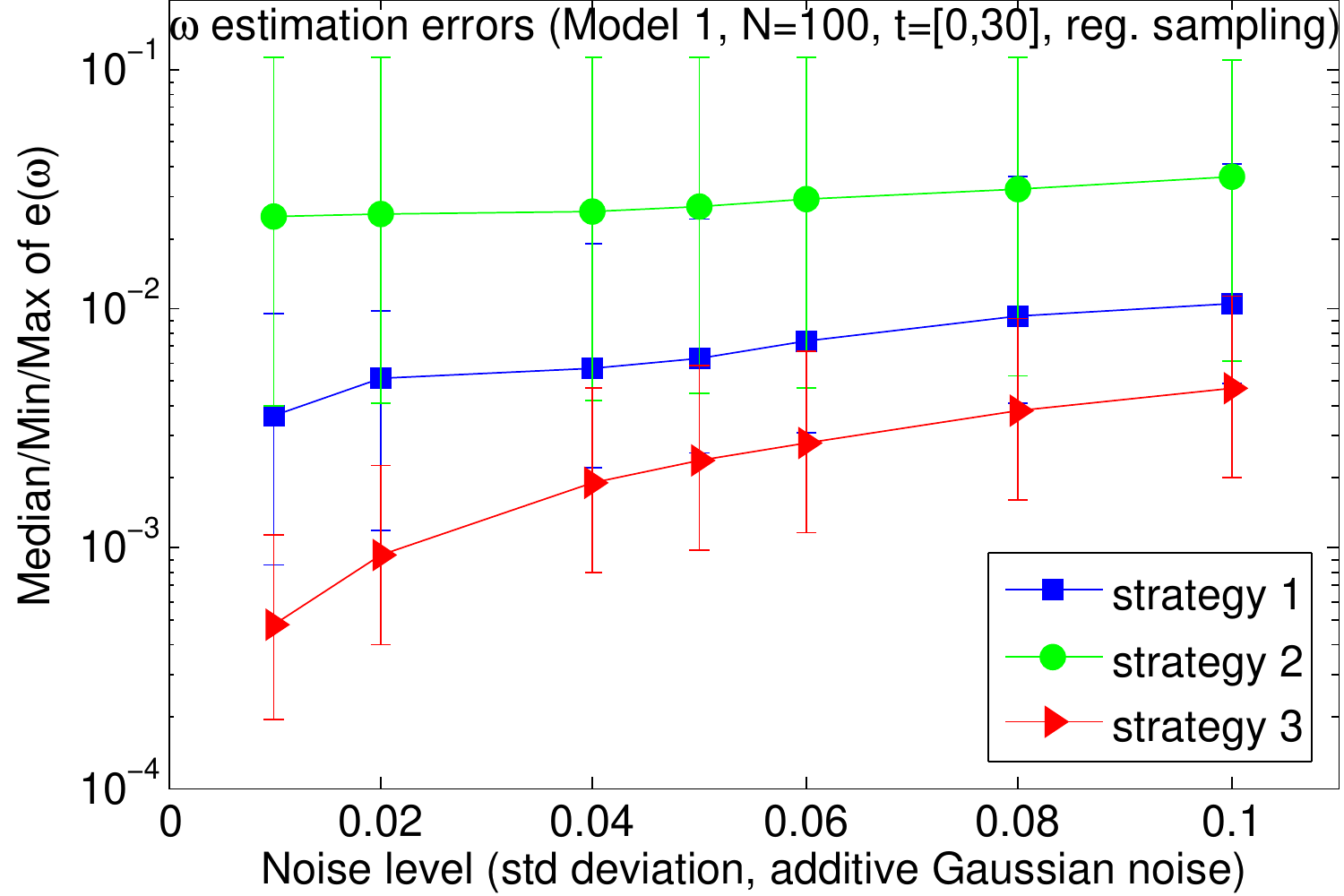} \hfill
\includegraphics[width=0.49\textwidth]{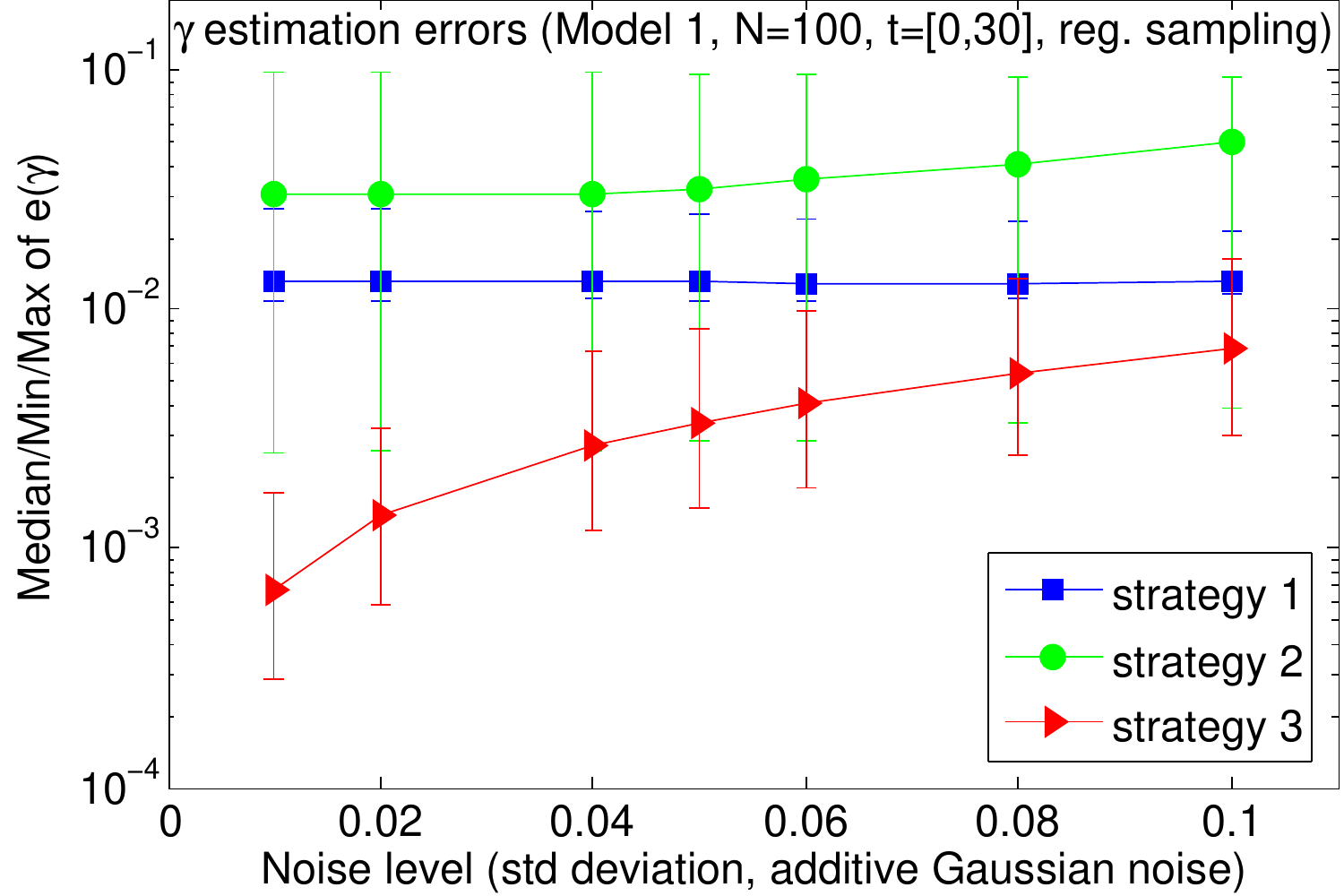}
\caption{Minimum, maximum and median of relative error (averaged over
  1000 runs for each system and noise level) of $\omega$ (left) and
  $\gamma$ estimates (right) as a function of the magnitude of the
  Gaussian noise for $10$ model systems (Table~\ref{table:models}).}
\label{fig:compare1}
\end{figure*}

\begin{figure*}
\subfigure[\ $\omega$ estimates: Strategy 1]
{\includegraphics[width=0.32\textwidth]{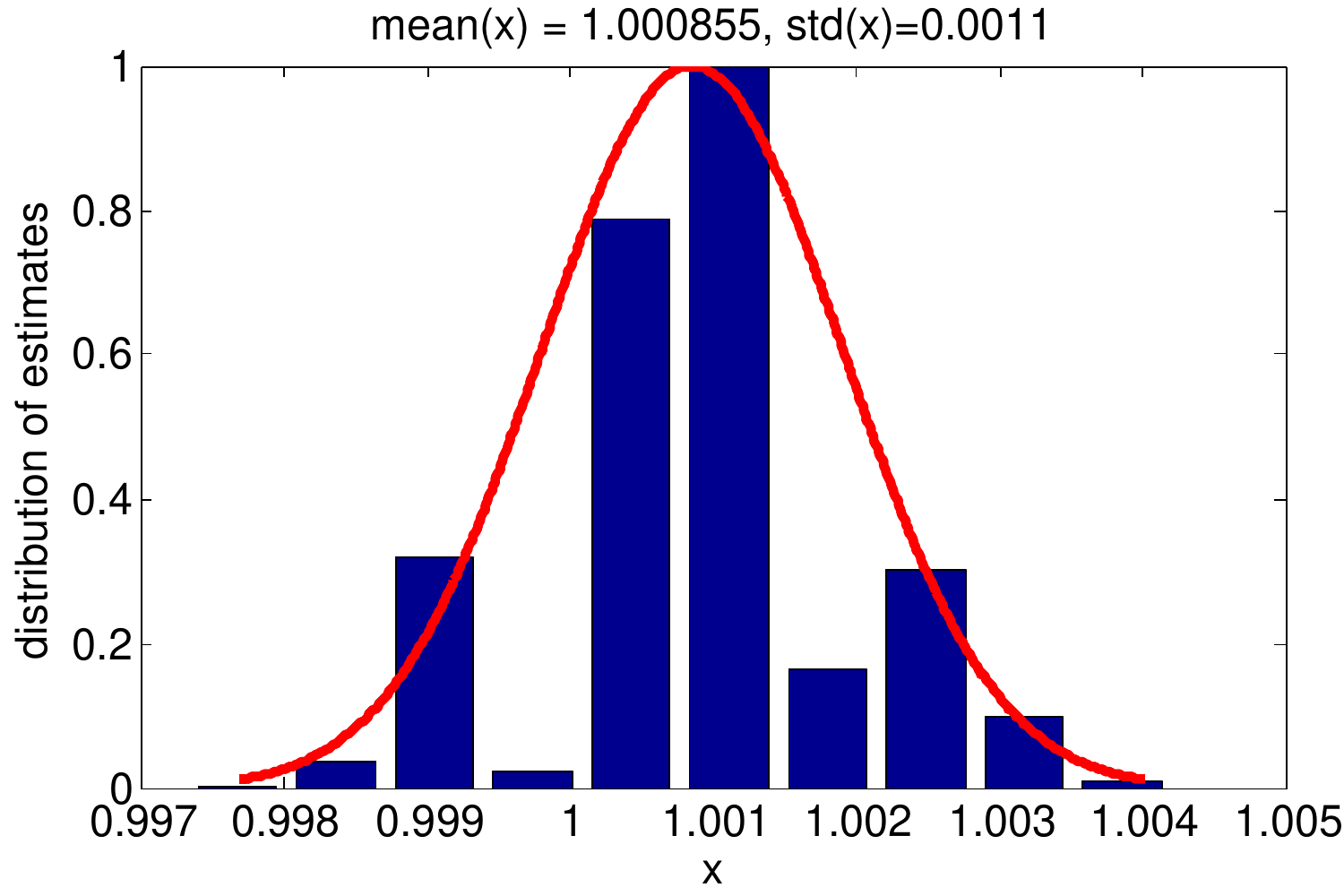}} \hfill
\subfigure[\ $\omega$ estimates: Strategy 2]
{\includegraphics[width=0.32\textwidth]{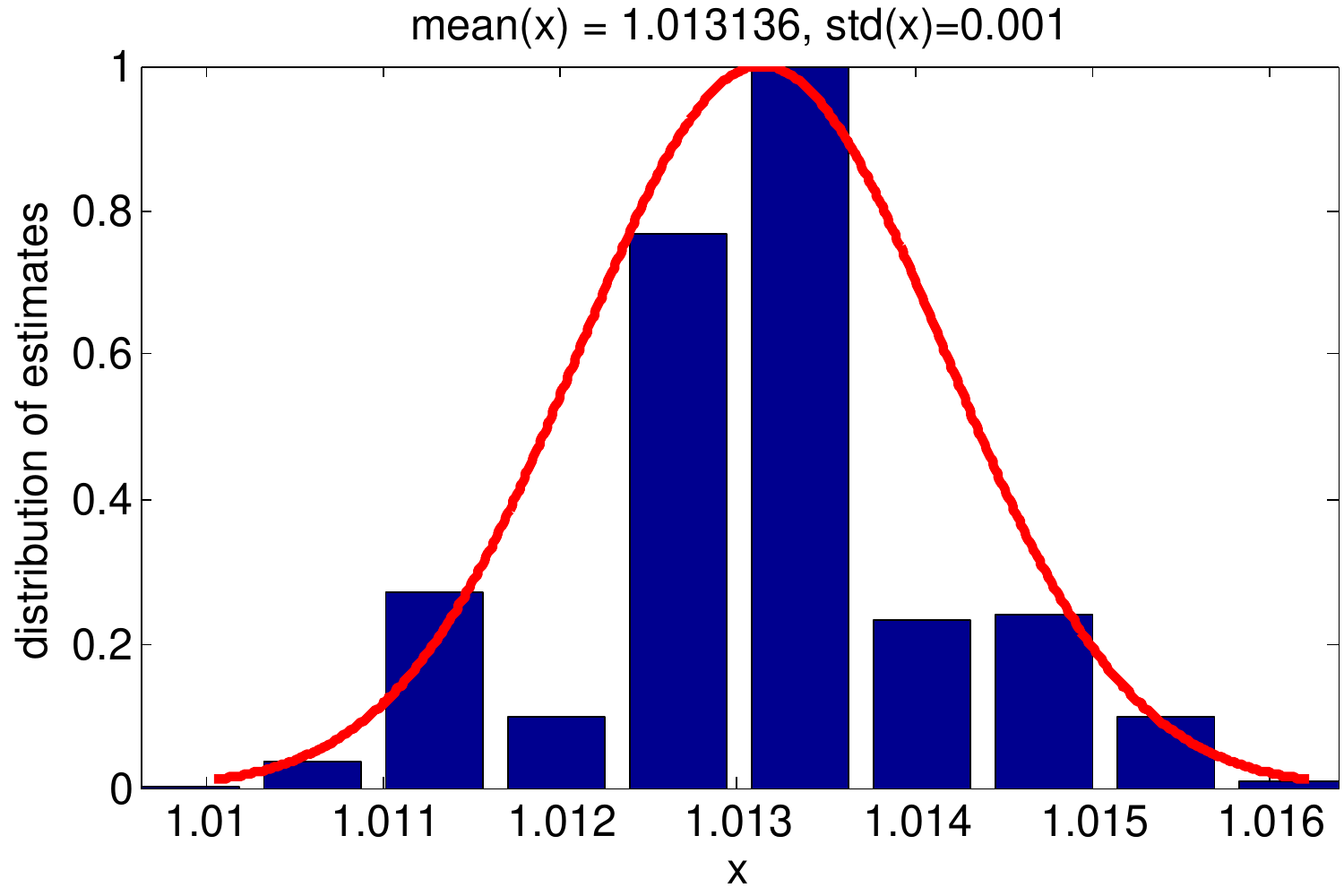}} \hfill
\subfigure[\ $\omega$ estimates: Strategy 3]
{\includegraphics[width=0.32\textwidth]{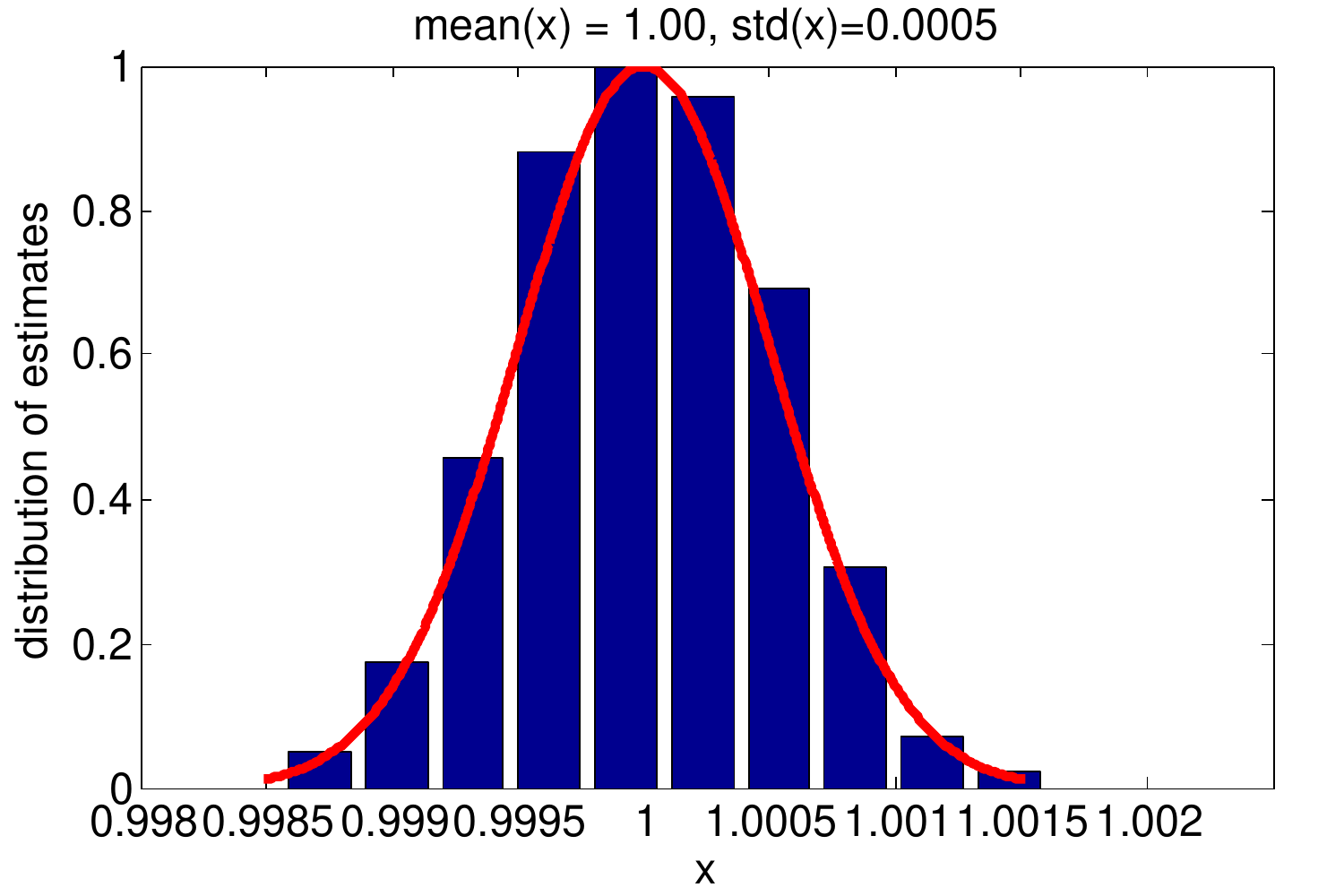}} \\
\subfigure[\ $\gamma$ estimates: Strategy 1]
{\includegraphics[width=0.32\textwidth]{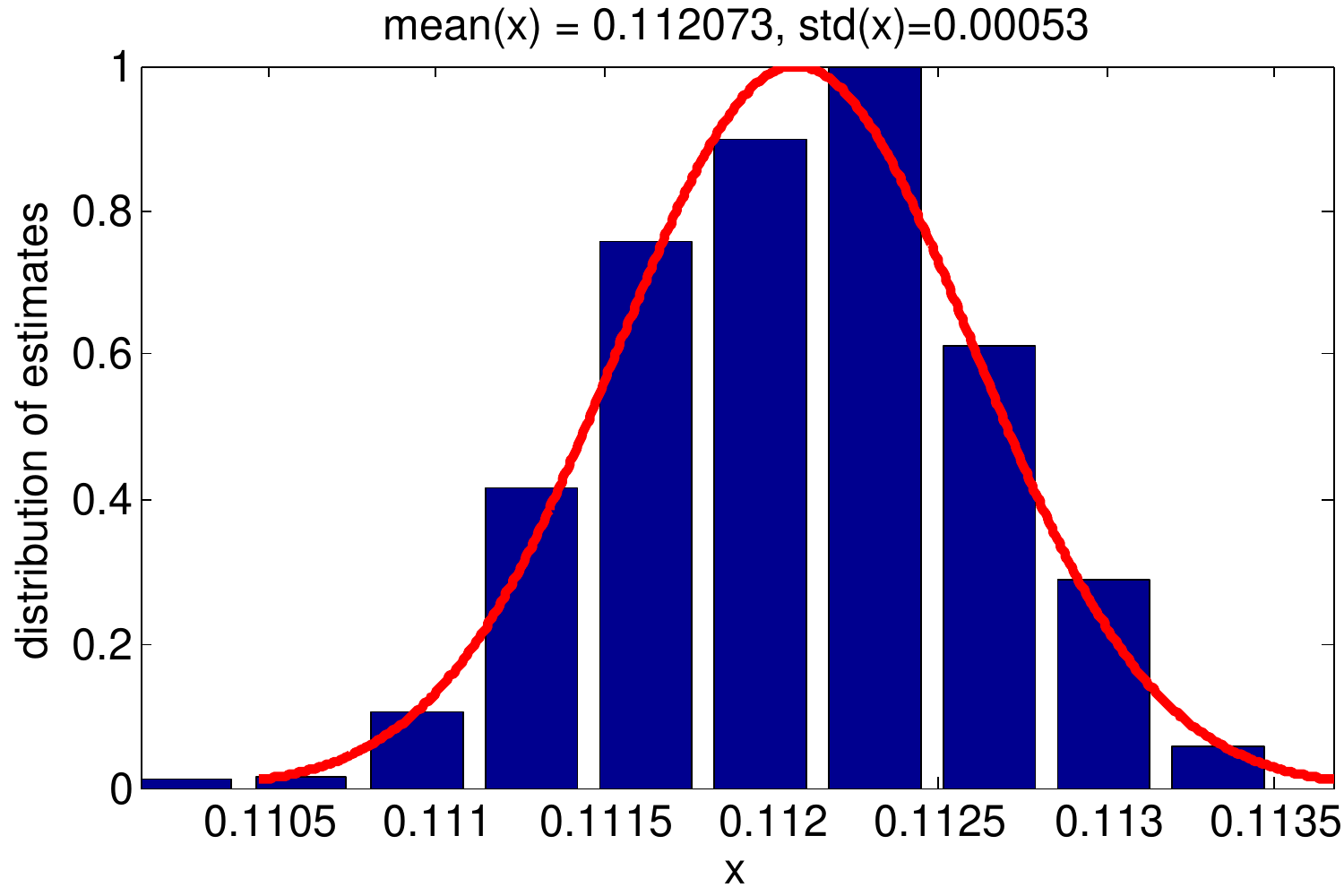}} \hfill
\subfigure[\ $\gamma$ estimates: Strategy 2]
{\includegraphics[width=0.32\textwidth]{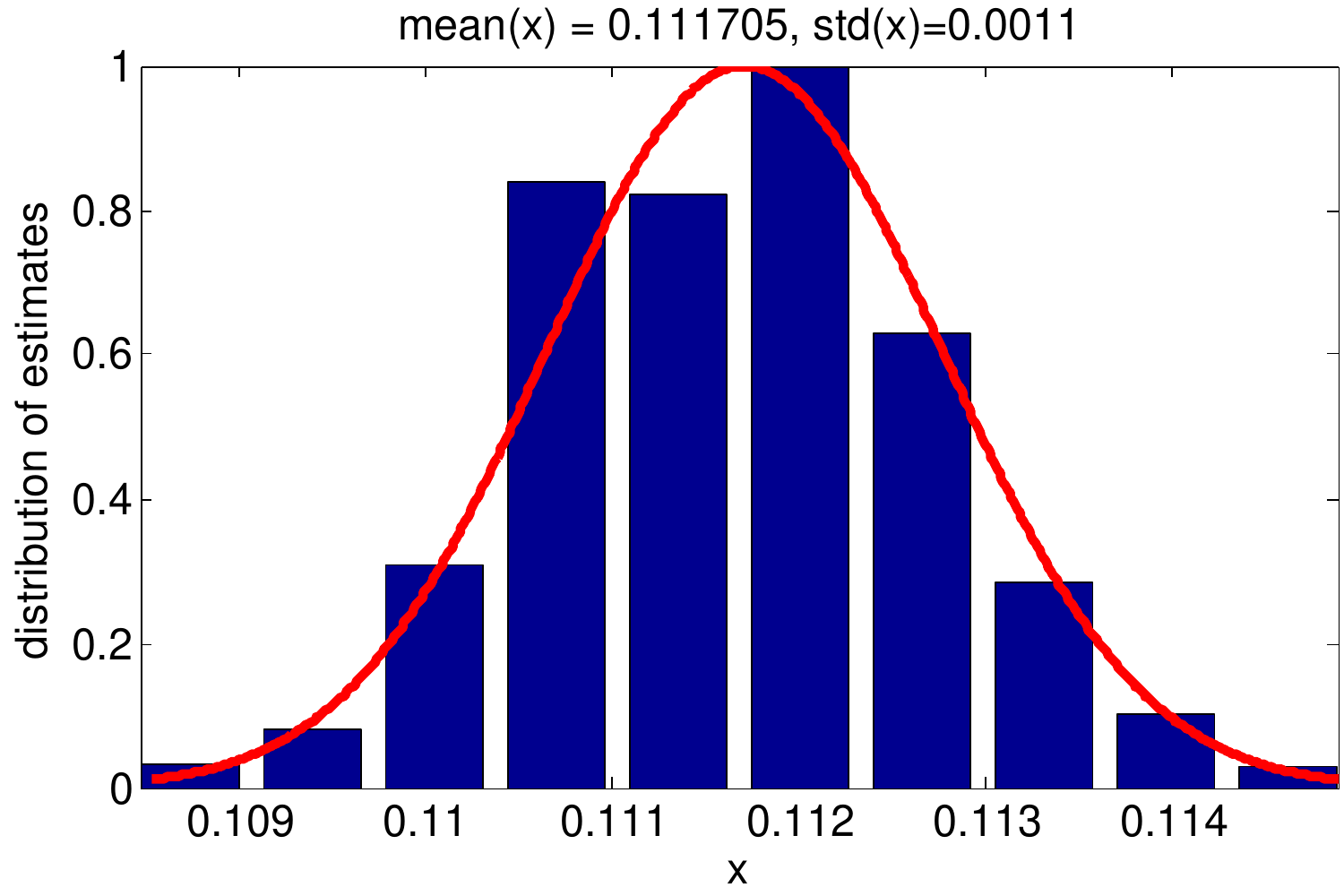}} \hfill
\subfigure[\ $\gamma$ estimates: Strategy 3]
{\includegraphics[width=0.32\textwidth]{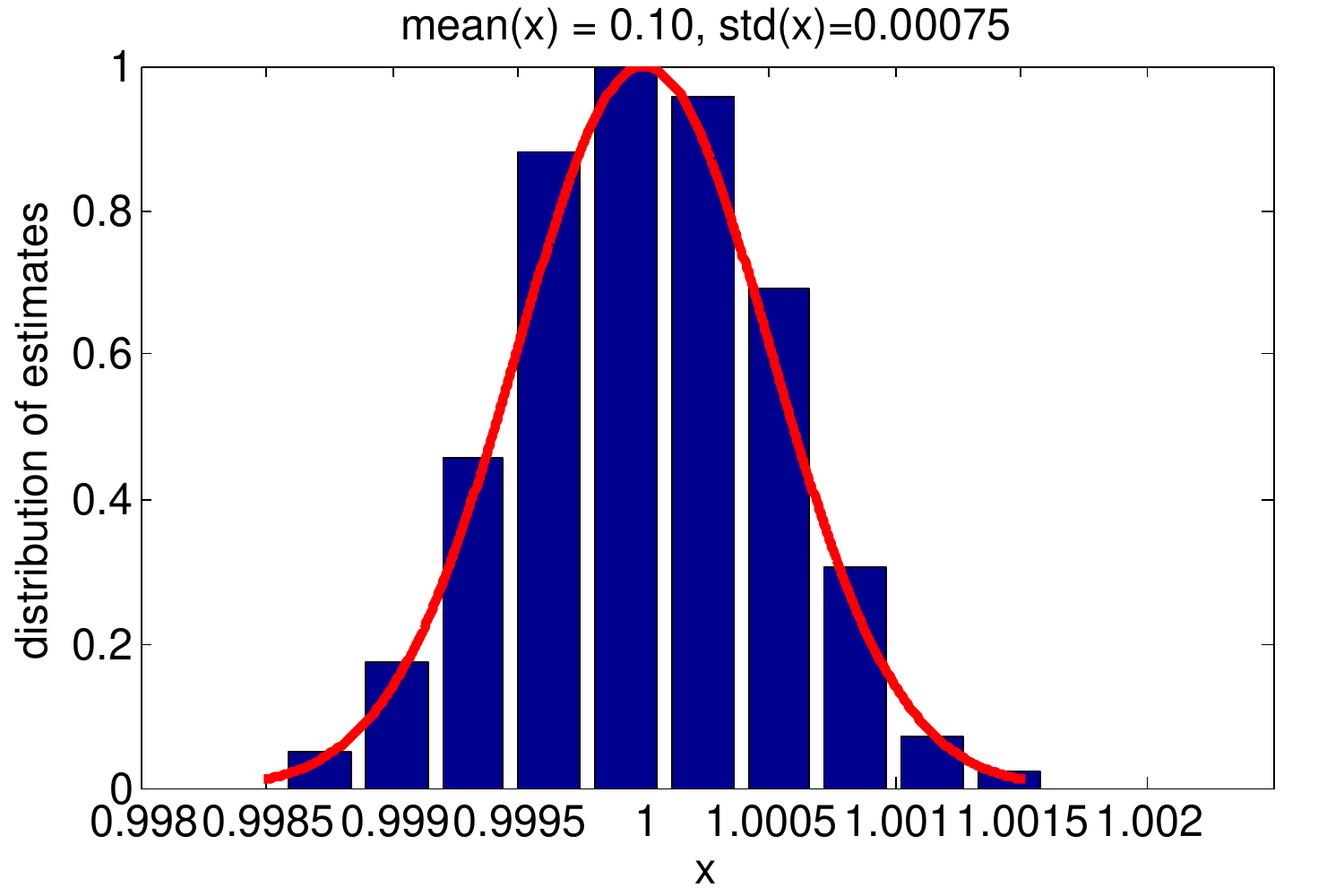}} \\
\caption{Distribution of $\omega$ and $\gamma$ estimates for 1000 runs for model 1 with
1\% Gaussian noise for strategies 1, 2 and 3.}
\label{fig:estimation-bias}
\end{figure*}

\begin{figure*}
\subfigure[\ 1\% Gaussian Noise]
{\includegraphics[width=0.32\textwidth]{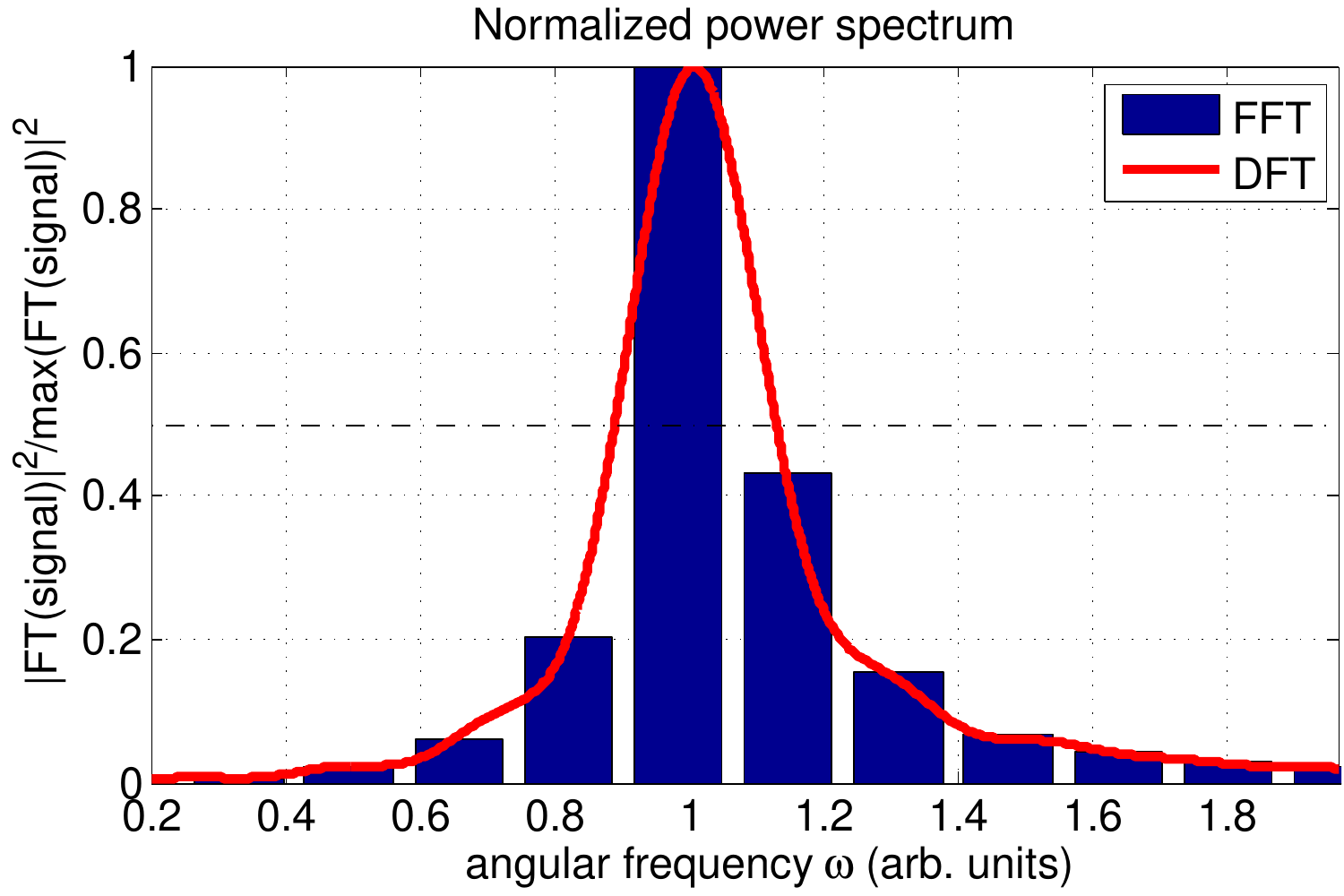}} \hfill
\subfigure[\ 5\% Gaussian Noise]
{\includegraphics[width=0.32\textwidth]{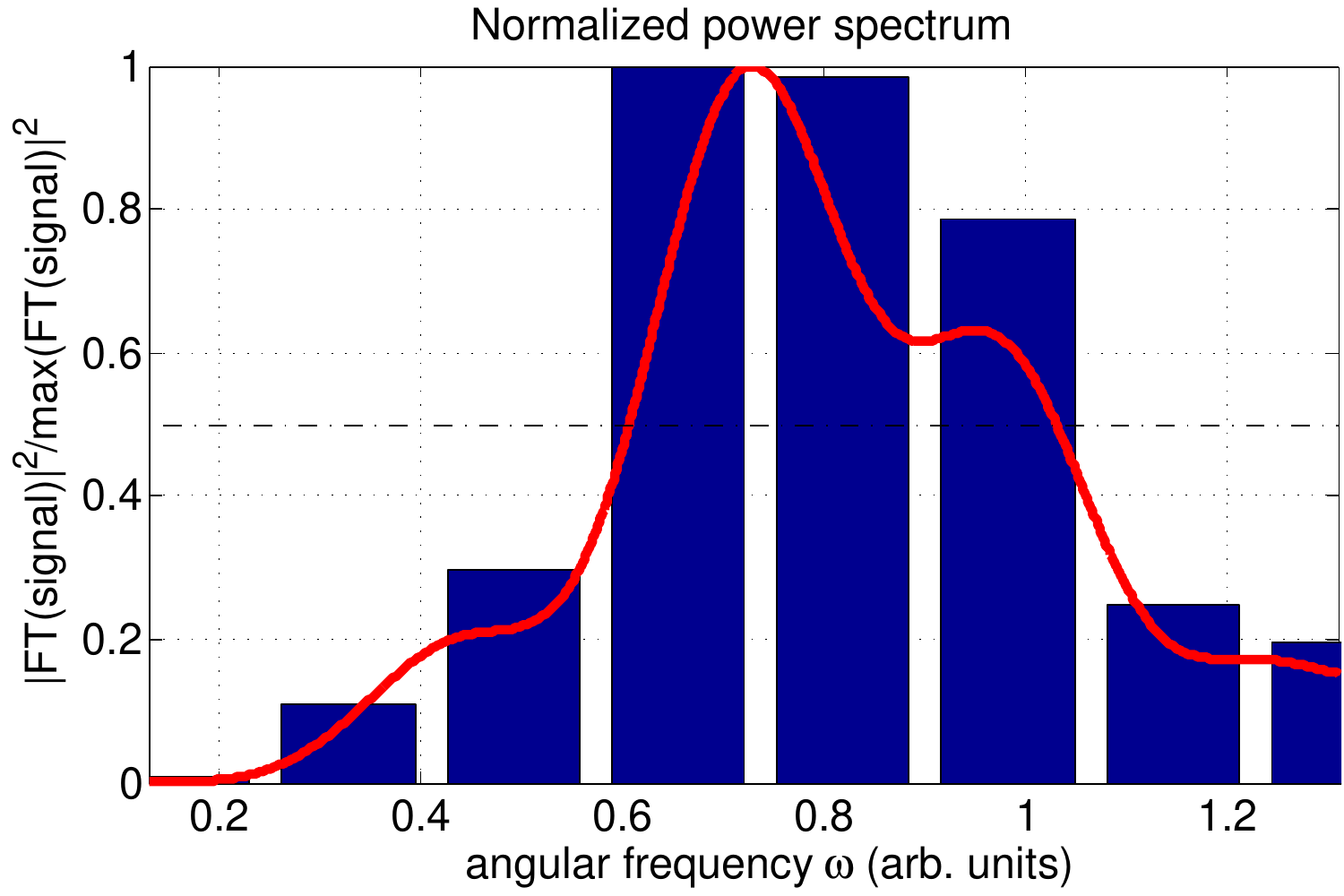}}\hfill
\subfigure[\ 10\% Gaussian Noise]
{\includegraphics[width=0.32\textwidth]{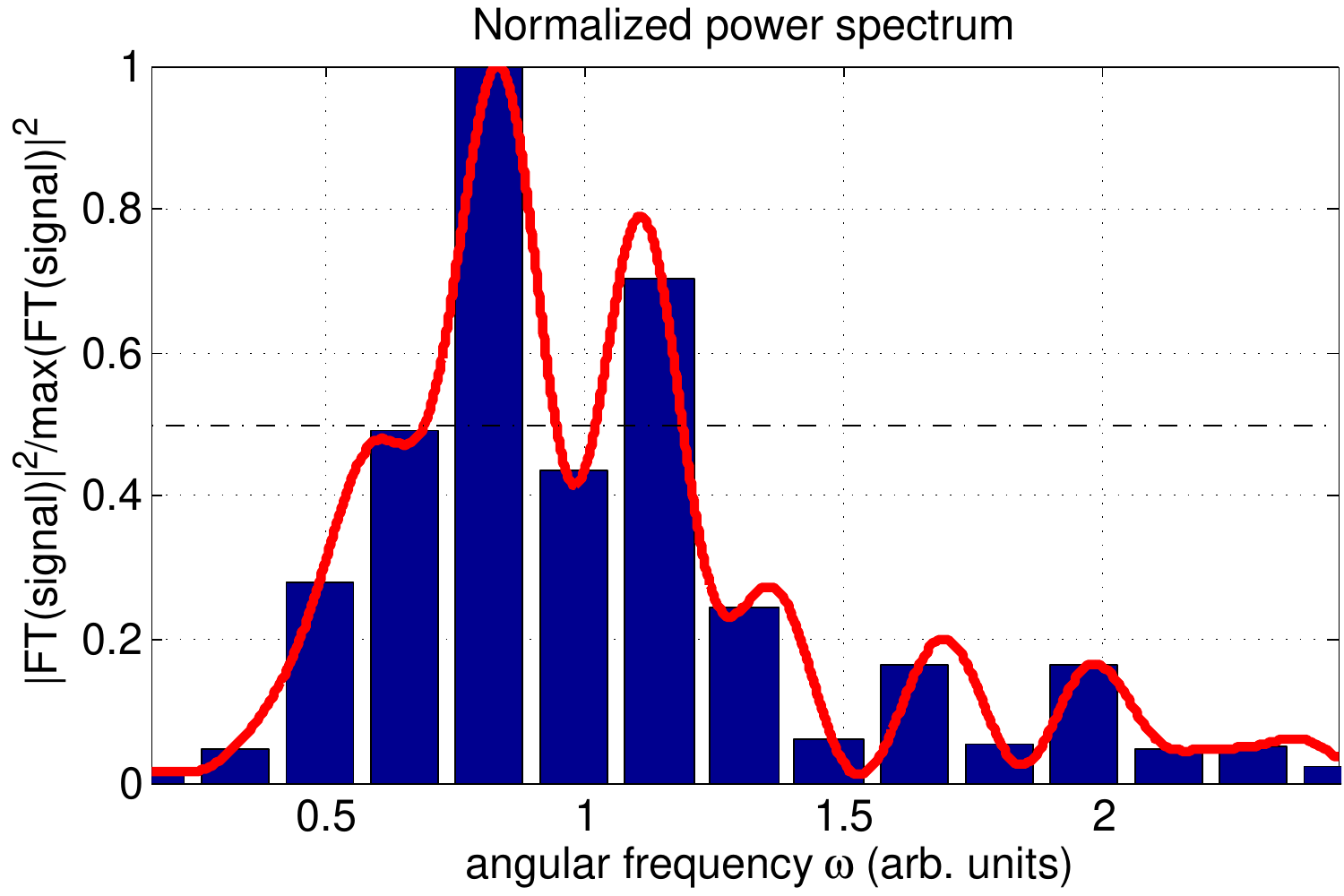}}
\caption{Limits of Fourier resolution and difficulty in estimating peak width for short,
noisy signals.}
\label{fig:fourier-problems}
\end{figure*}

\begin{figure}
\includegraphics[width=0.49\textwidth]{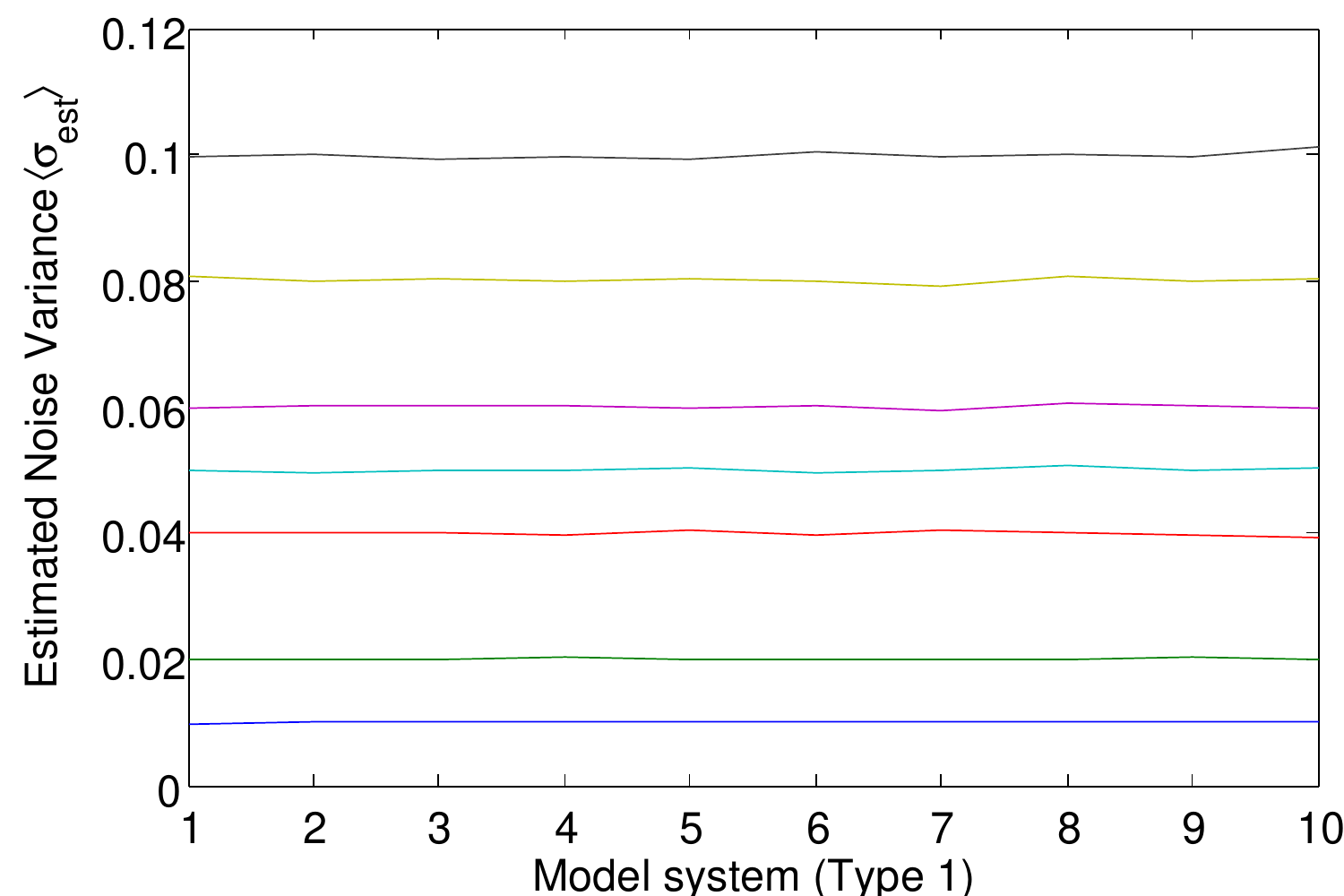}
\caption{The estimated noise level $\sigma$ of the measurement data
  for 10 model systems of type 1 obtained from Strategy 3 closely
  track the actual noise levels of the simulated data $(0.01, 0.02,
  0.04, 0.05, 0.6, 0.8, 1.0)$.}
\label{fig:compare1c}
\end{figure}

\subsection{Ensemble measurements with Gaussian noise}

To compare the different estimation strategies for discretely sampled
signals with Gaussian noise we simulate the measurement result $d_k$
at time $t_k$. The expected signal $p(t_k)$ was calculated based on
the selected model and Gaussian noise of mean $0$ and standard
deviation $\sigma$ added to each value. Fig.~\ref{fig:signal} (left)
shows an example of an ideal measurement signal and simulated data
with uniform sampling at times $t_n = n\Delta t$ with $\Delta t=0.3$.

Fig.~\ref{fig:compare1} compares the errors according to (\ref{eq:error})
for the three strategies. Strategy~2, probably the most
common technique for estimating the frequency and dephasing parameter
using the position and width of the peak in the Fourier spectrum,
actually gives the least accurate and least precise estimates --- the
median error of the estimated values is large, as is the spread of the
errors for different systems as indicated by the large error bars.
Strategy~1 produces slightly improved estimates, but parameter
estimates based on Strategy 3 are significantly better.  The results
are similar for $\omega$ and $\gamma$.  Fig.~\ref{fig:estimation-bias}
furthermore suggests that Strategies~1 and 2 are \emph{not} unbiased
estimators. The mean of the distribution over the estimation runs does
not appear to converge to the true value of the parameter even for
very lowest noise level and 1000 runs. Strategy~3, however, appears to
be an unbiased Gaussian estimator.

One interesting feature of Strategies~1 and 2 is that the median
estimation errors appear to be almost constant over the range of noise
levels considered, while for Strategy~3 the error increases with
increasing noise level, as one would expect. A probable reason for
this is that the uncertainties in the position, and indirectly the
width, of the peaks in the Fourier spectrum primarily depend on the
length of the signal $T$.  Specifically, for a fixed number of
samples,~\cite{Cole2006} found that the uncertainty in the parameter
estimates was mainly proportional to $1/\sqrt{T}$.  This would explain
why the accuracy of the estimates obtained from the Fourier-based
strategies appears roughly constant as the signal length and number of
samples were both fixed in our simulated experiments ($T=30$,
$N_t=100$).  So it might be argued that the Fourier-based strategies
are less sensitive to noise. However, it is important to notice that
even for noise with $\sigma=0.1$, Strategy~3 still outperforms the
other strategies in all cases.

Furthermore, accurately and precisely estimating location and width of
a peak in the Fourier spectrum for a relatively short, noisy signal
can be challenging, as illustrated by the power spectrum examples in
Fig.~\ref{fig:fourier-problems}.  The blue bars show the $|F(k)|^2$,
where $F(k)$ is the discrete Fourier transform of the measured
discrete signal
\begin{equation}
 \label{eq:DFT}
  F(k) = \sum_{n=1}^{N_t}  d_n' e^{-2\pi i(k-1)(n-1)/N_t}, \quad 1 \le k \le N_t,
\end{equation}
computed using the Fast Fourier Transform (FFT), after
centering and rescaling, $\vec{d}' = (\vec{d}-\bar{d})/d_{\max}$ with
$\bar{d}=\tfrac{1}{N_t} \sum_{n=1}^{N_t} d_n$ and
$d_{\max}= \max |d_n-\bar{d}|$.  The red curve is an approximation to the
continuous Fourier transform
\begin{equation}
\label{eq:FT}
  F(\omega) = \int_{-\infty}^\infty \!\!\! f(t) e^{-i\omega t} dt
            \approx \sum_{n=1}^{N_t} d_n' e^{i\omega t_n} \tfrac{1}{2} (\Delta t_n + \Delta t_{n-1})
\end{equation}
where the integral has been approximated using the trapezoidal rule
with $\Delta t_n = t_{n+1}-t_n = T/N_t$ for $n=1,\ldots,N_t-1$ and
$\Delta t_0=\Delta t_{N_t}=0$.  The left figure shows a ``good'' power
spectrum for a low-noise input signal.  Even in this case the
frequency resolution is limited but the peak has a more or less
Lorentzian shape and the width is well defined.  However, for
increasing noise the peak can become increasingly distorted (center)
and for very noisy signals it may even become split (right) making
width estimation difficult and assumptions about kurtosis and skewness
are no longer valid.

A further advantage of Strategy~3 is that it also provides direct
estimates for the noise variance~\cite{Bretthorst}
\begin{equation}
  \sigma = \tfrac{1}{N_t-m_b-2} (N_t \ave{\vec{d}^2} - m_b \ave{\vec{h}^2})
\end{equation}
and Fig.~\ref{fig:compare1c}
shows that the estimates are very accurate across the board.

\begin{figure*}
\includegraphics[width=0.49\textwidth]{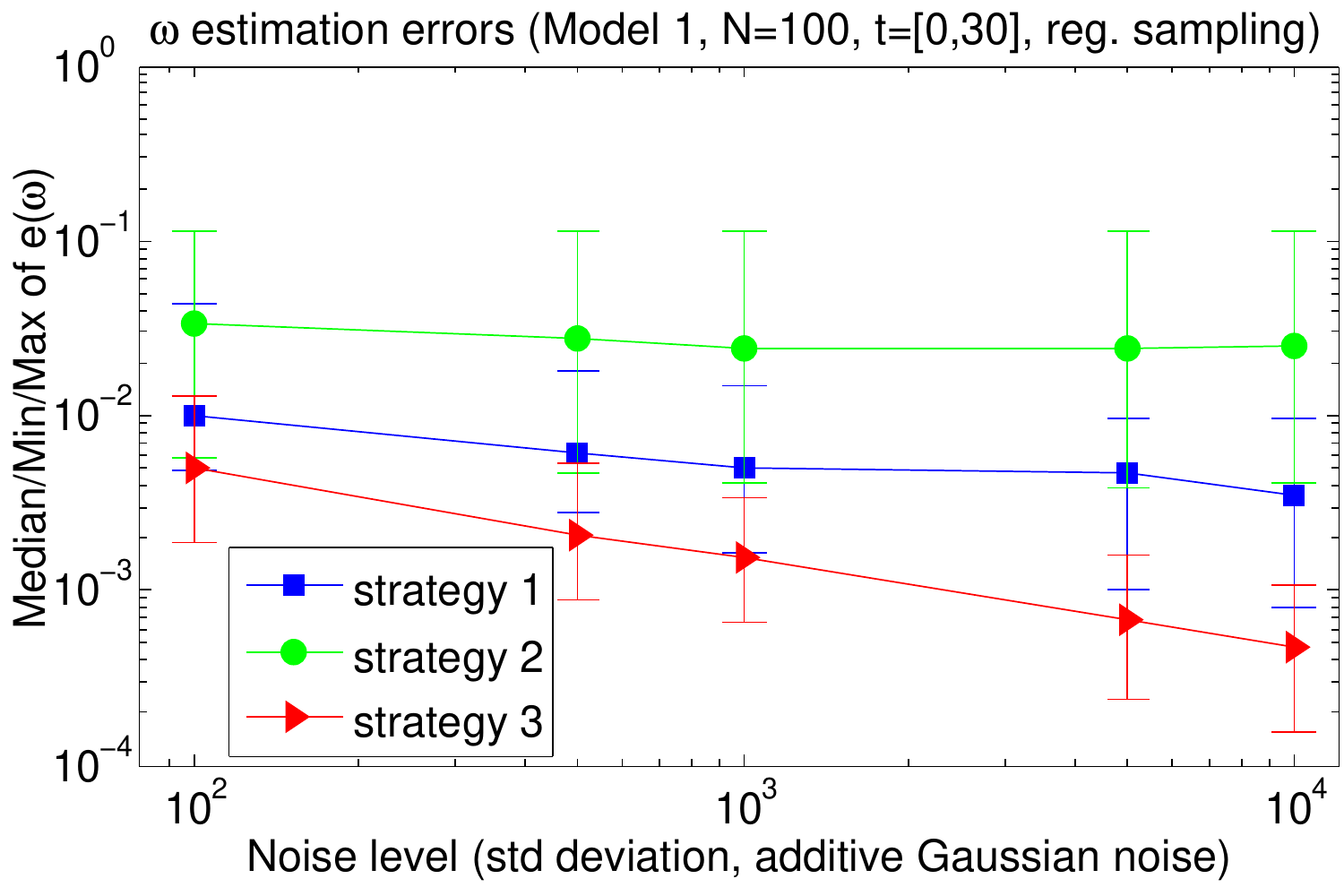} \hfill
\includegraphics[width=0.49\textwidth]{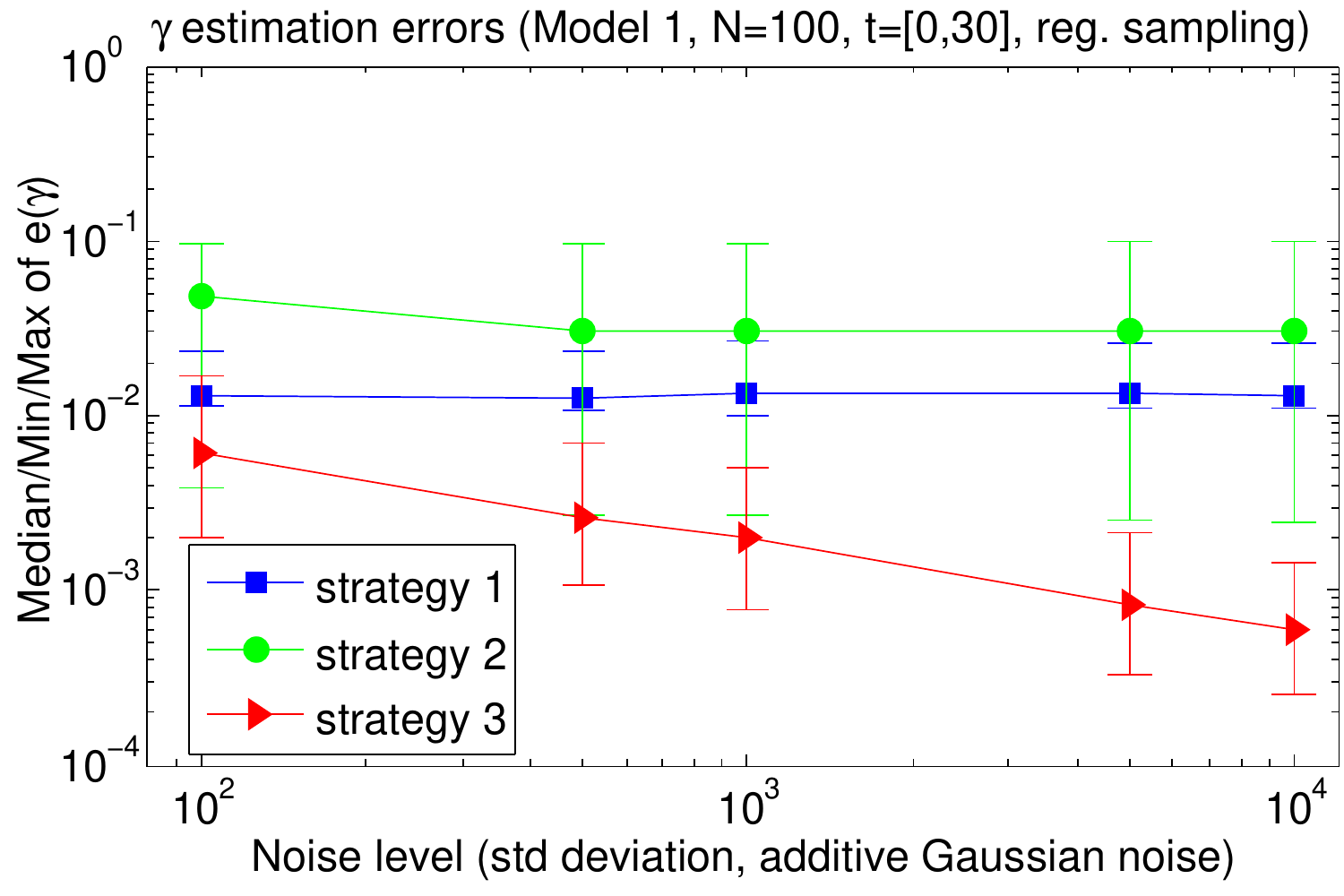}
\caption{Minimum, maximum and median of relative error of $\omega$
  (left) and $\gamma$ estimates (right) as a function of the number of
  single-shot measurement repetitions per data point, $N_e$, for $10$
  model systems (Table~\ref{table:models}).}
\label{fig:compare2}
\end{figure*}

\begin{figure}
\includegraphics[width=0.49\textwidth]{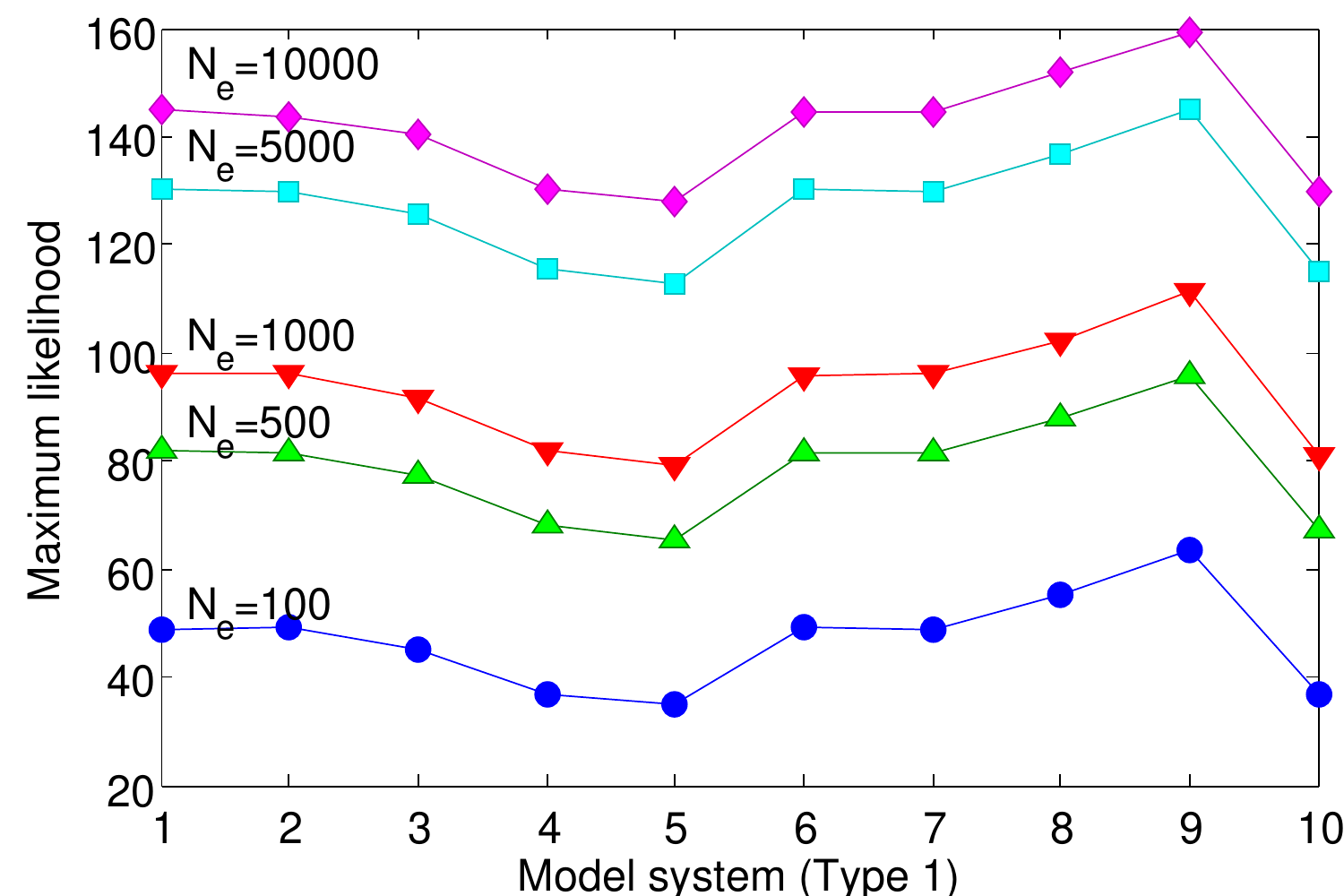}
\caption{Maximum likelihood for 10 model systems, averaged
         over 100 runs each, obtained from Strategy 3.}
\label{fig:compare2b}
\end{figure}

\begin{figure}
\includegraphics[width=0.49\textwidth]{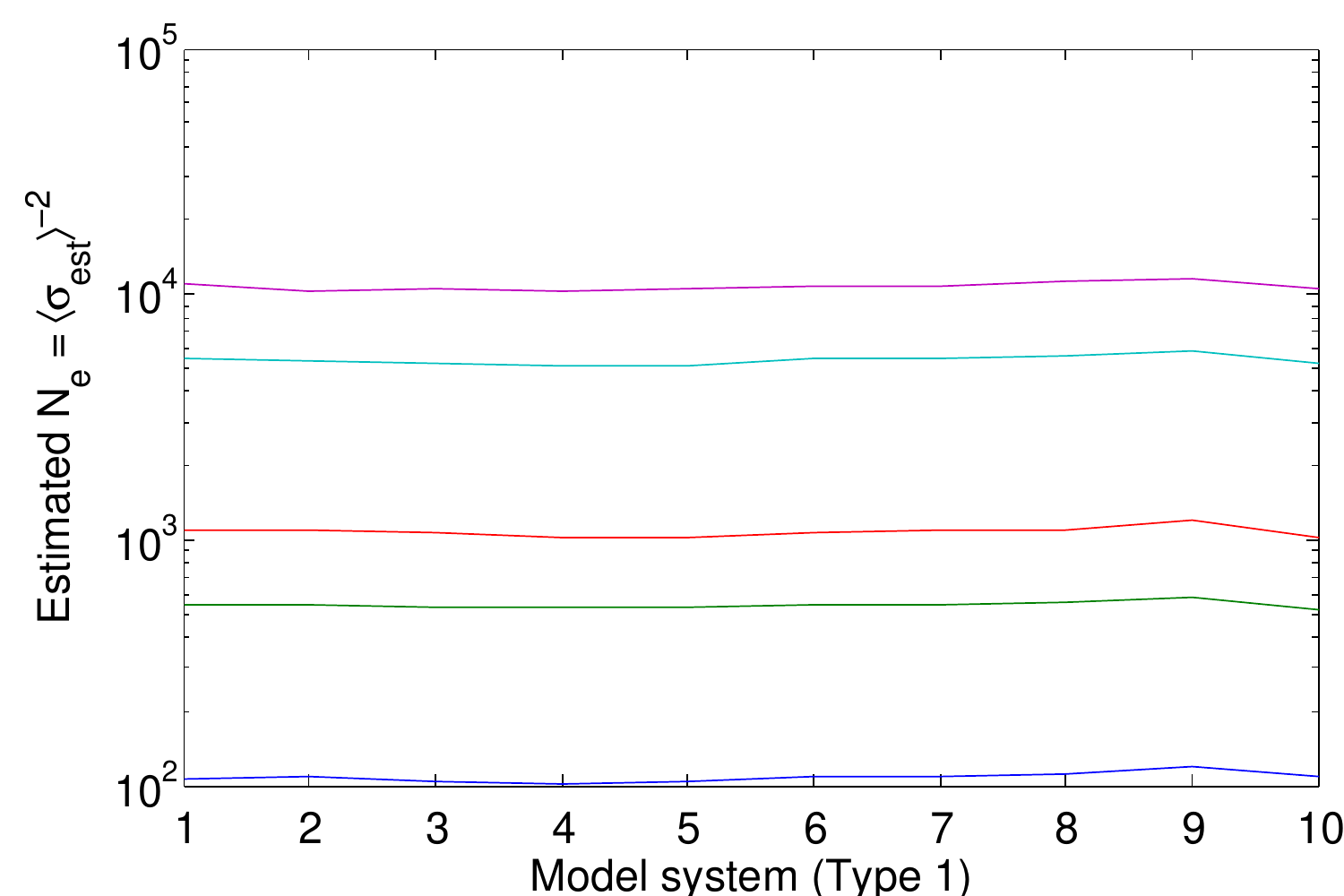}
\caption{Estimated $N_e = \ave{\sigma_{\rm est}}^{-2}$ for single shot
  measurements for 10 model systems, averaged over 100 runs each,
  obtained from Strategy 3.  The $N_e$ estimates closely track the
  actual number of repetitions of the single shot measurements for the
  simulated data $(100, 500, 1000, 5000, 10000)$.}
\label{fig:compare2c}
\end{figure}

\subsection{Single-system measurements}

To assess if there are significant differences in the performance of
different estimation strategies in the presence of projection noise,
we repeat the analysis in the previous subsection for the same 10
model systems, sampled over the same time interval $[0,30]$, but with
various levels of projection noise added instead of Gaussian noise.
Fig.~\ref{fig:signal} (right) shows an example of an ideal measurement
signal and simulated data. Fig.~\ref{fig:compare2} shows the relative
errors for the different estimation strategies for the same model
systems but subject to (simulated) projection noise.  Strategy~3 again
performs significantly better than the other strategies.
Fig.~\ref{fig:compare2b} shows that the likelihood of the estimates
increases with increasing number of repetitions $N_e$, as expected. It
also shows again that the maximum likelihood for some model systems is
consistently higher than for others, as was observed for Gaussian
noise.

Fig.~\ref{fig:compare2c} shows that even the estimates for the noise
variance $\sigma^2$ obtained automatically with Strategy~3 are very
accurate in that the results obtained closely track the theoretical
values $\sigma^2 = 1/N_e$ expected for projection noise.

Overall this shows that although the noise strictly follows a Poisson
distribution in this case, we still obtain very good estimates of the
noise level for typical values of $N_e$ using a Gaussian error model
in the derivation of the maximum likelihood estimation strategy.  So
overall Strategy~3 appears to be consistently better than Strategies~1
and 2, independent of the types of measurements and their associated
noise for the two-level frequency and dephasing estimation problem.

\begin{figure}
\includegraphics[width=0.49\textwidth]{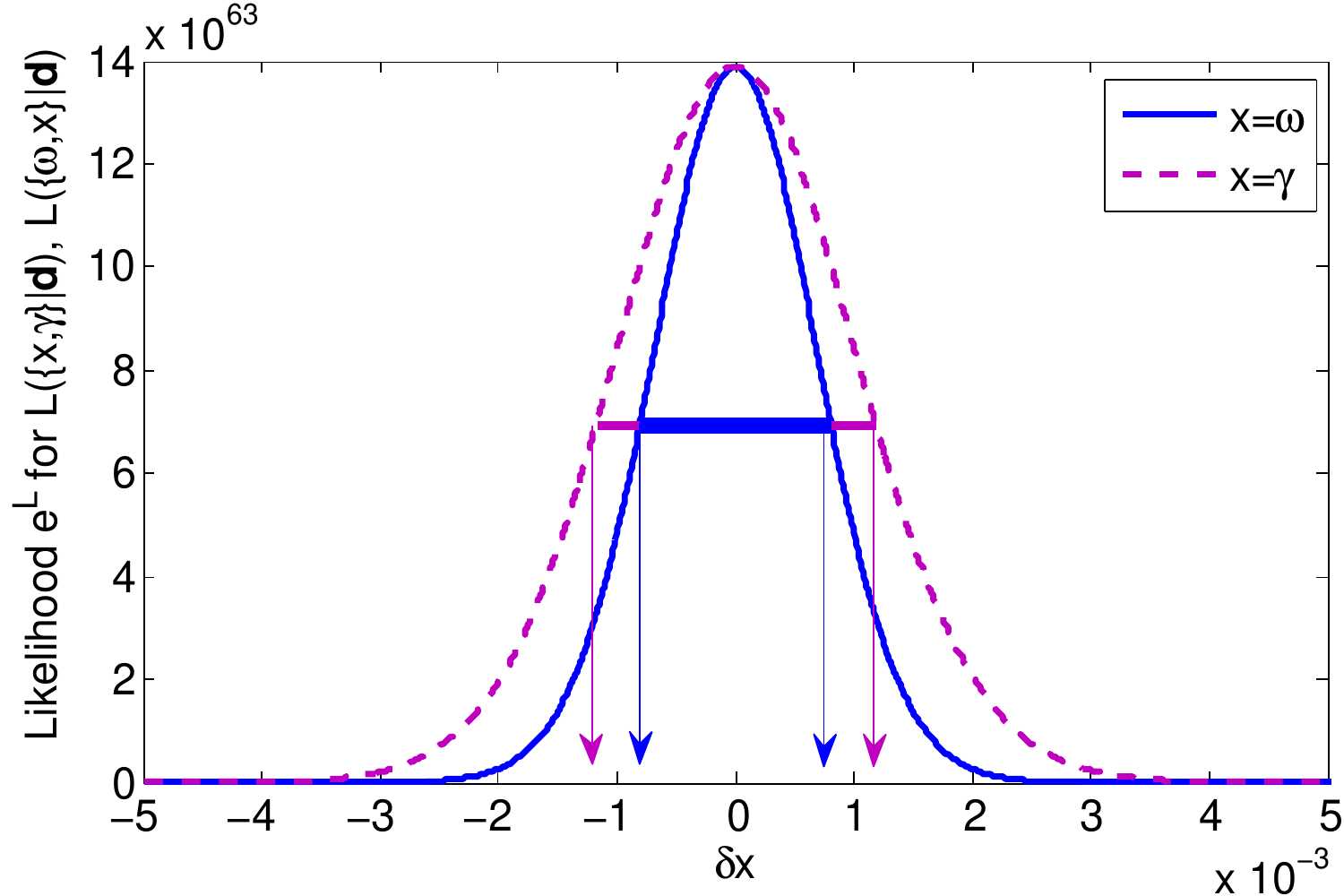}
\caption{Estimation of width of likelihood peak with regard to $\omega$ and $\gamma$.}
\label{fig:peak-width}
\end{figure}

\begin{figure*}
\includegraphics[width=0.49\textwidth]{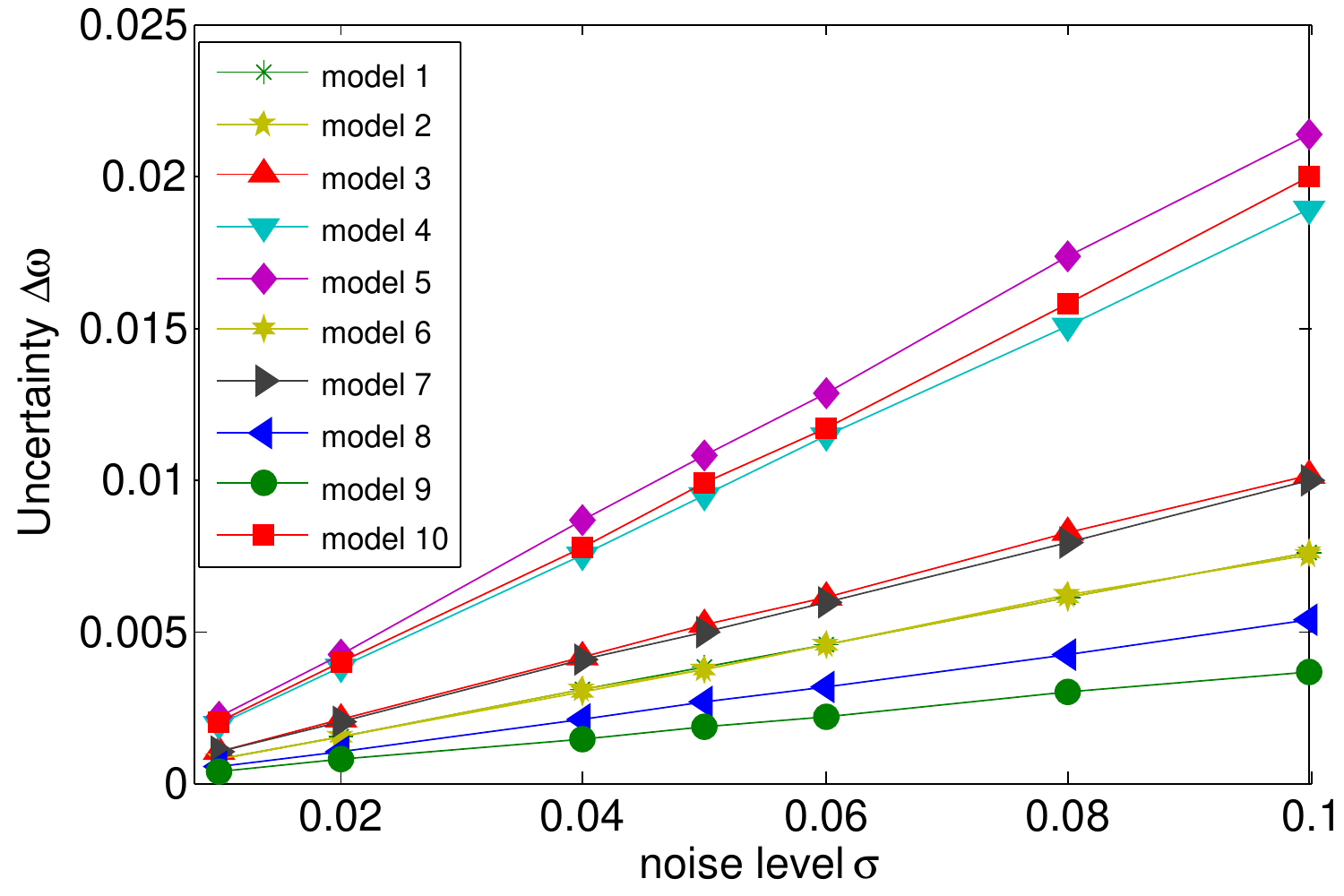} \hfill
\includegraphics[width=0.49\textwidth]{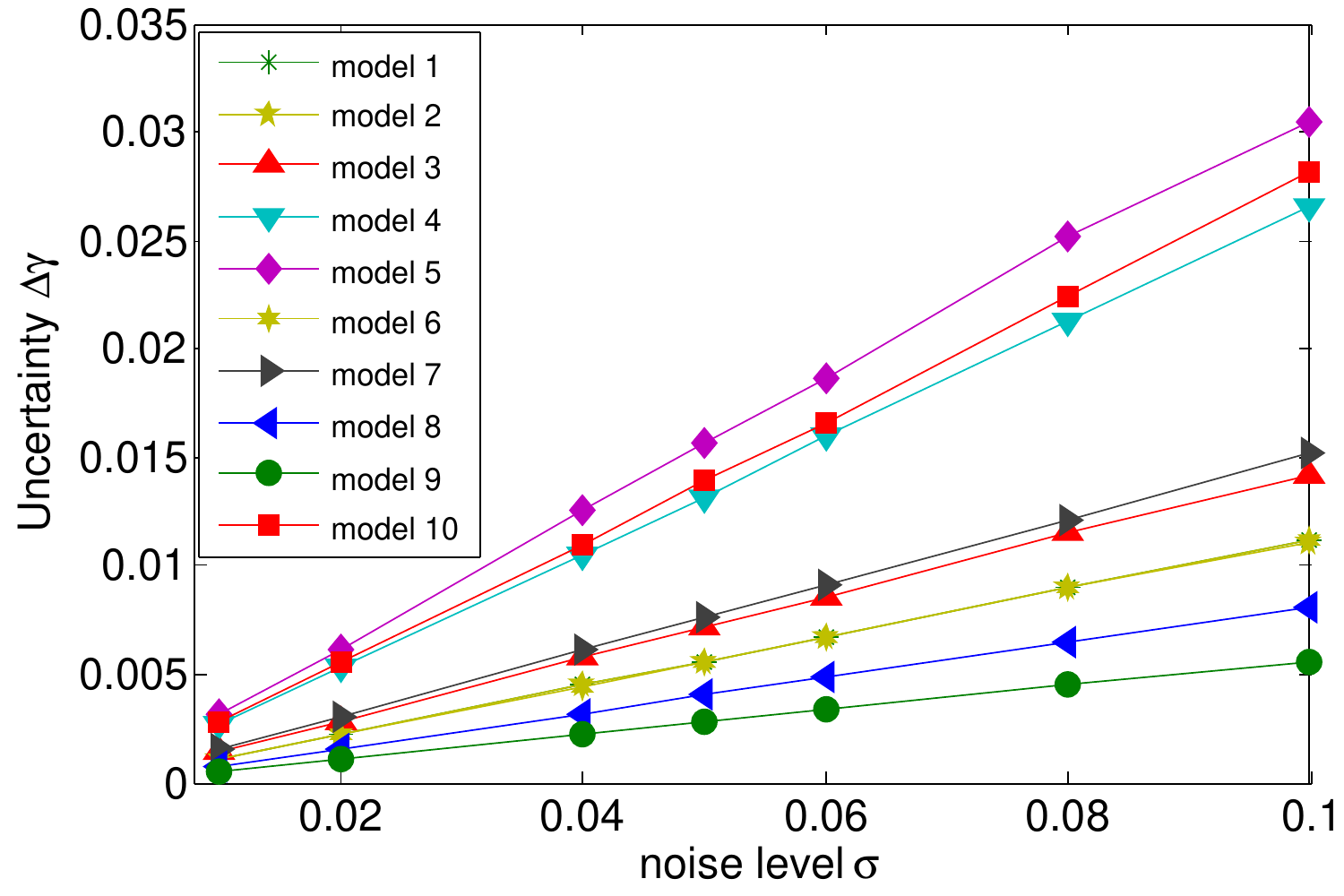}
\caption{Uncertainties of $\omega$ (left) and $\gamma$ estimates
  (right) for 10 model systems (Table~\ref{table:models}) as a
  function of Gaussian noise level.}
\label{fig:compare1a}
\end{figure*}

\begin{figure*}
\includegraphics[width=0.49\textwidth]{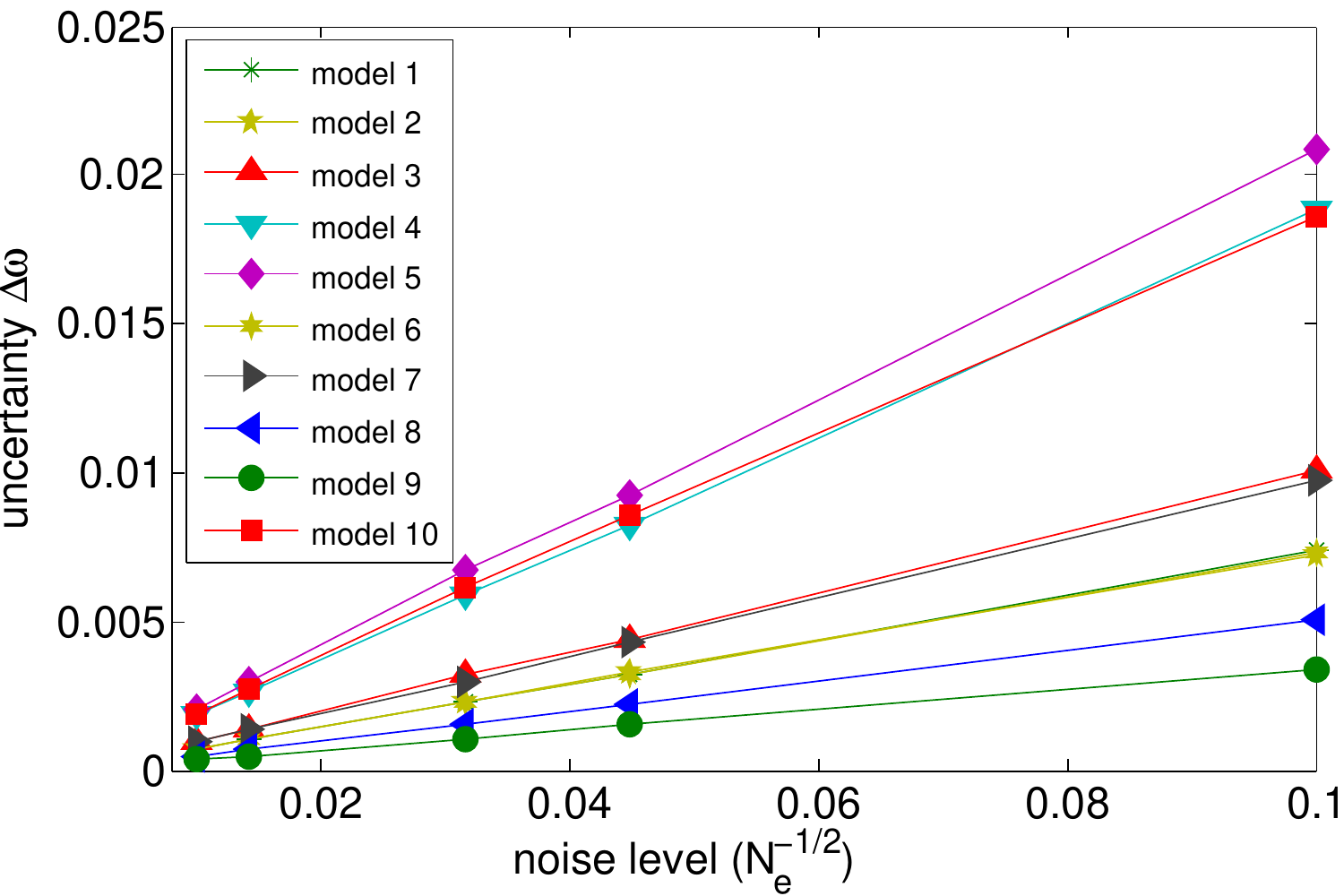} \hfill
\includegraphics[width=0.49\textwidth]{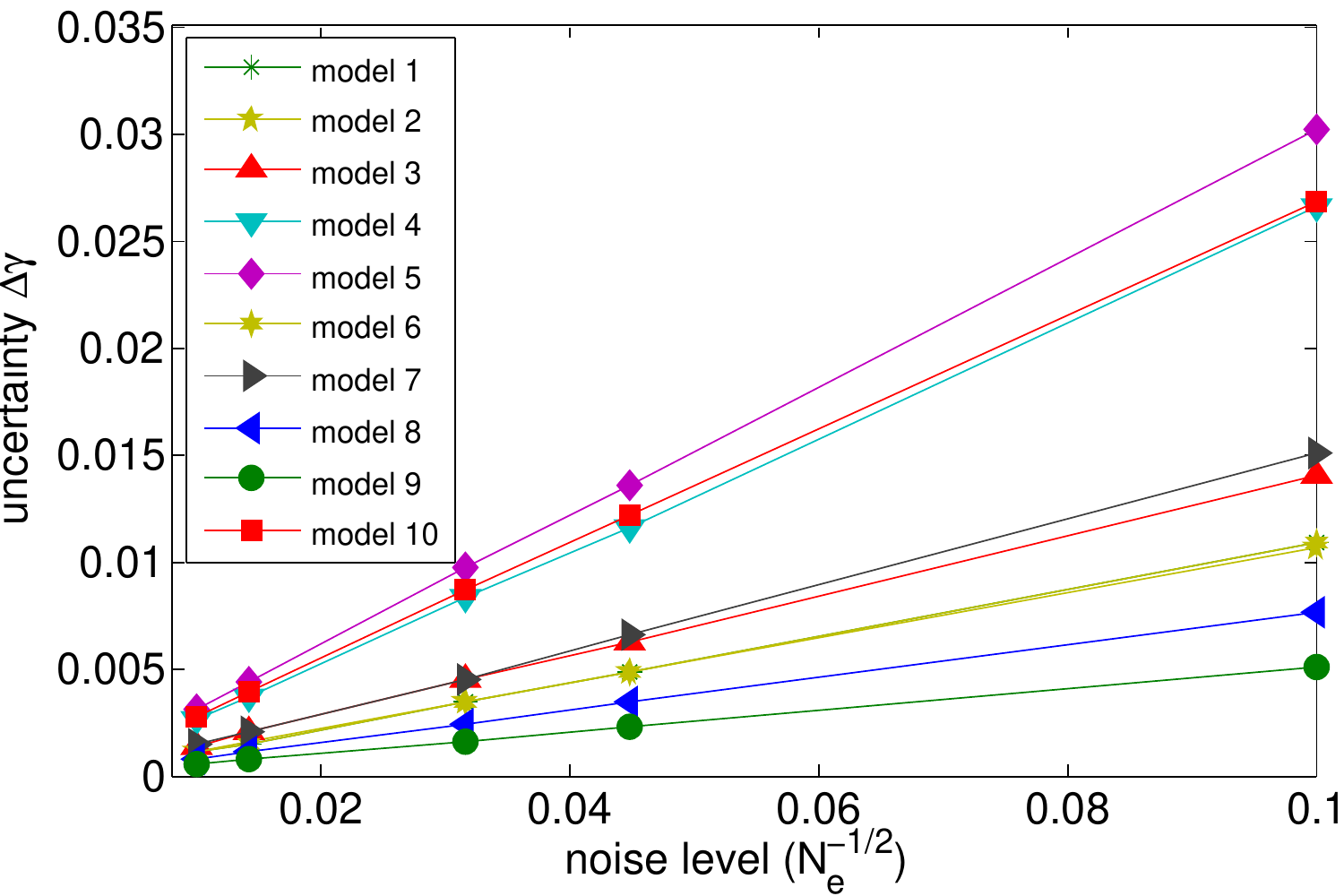} \\
\caption{Uncertainties of $\omega$ (left) and $\gamma$ estimates
  (right) for 10 model systems as a function of projection noise level.}
\label{fig:compare2a}
\end{figure*}

\begin{figure}
\includegraphics[width=0.49\textwidth]{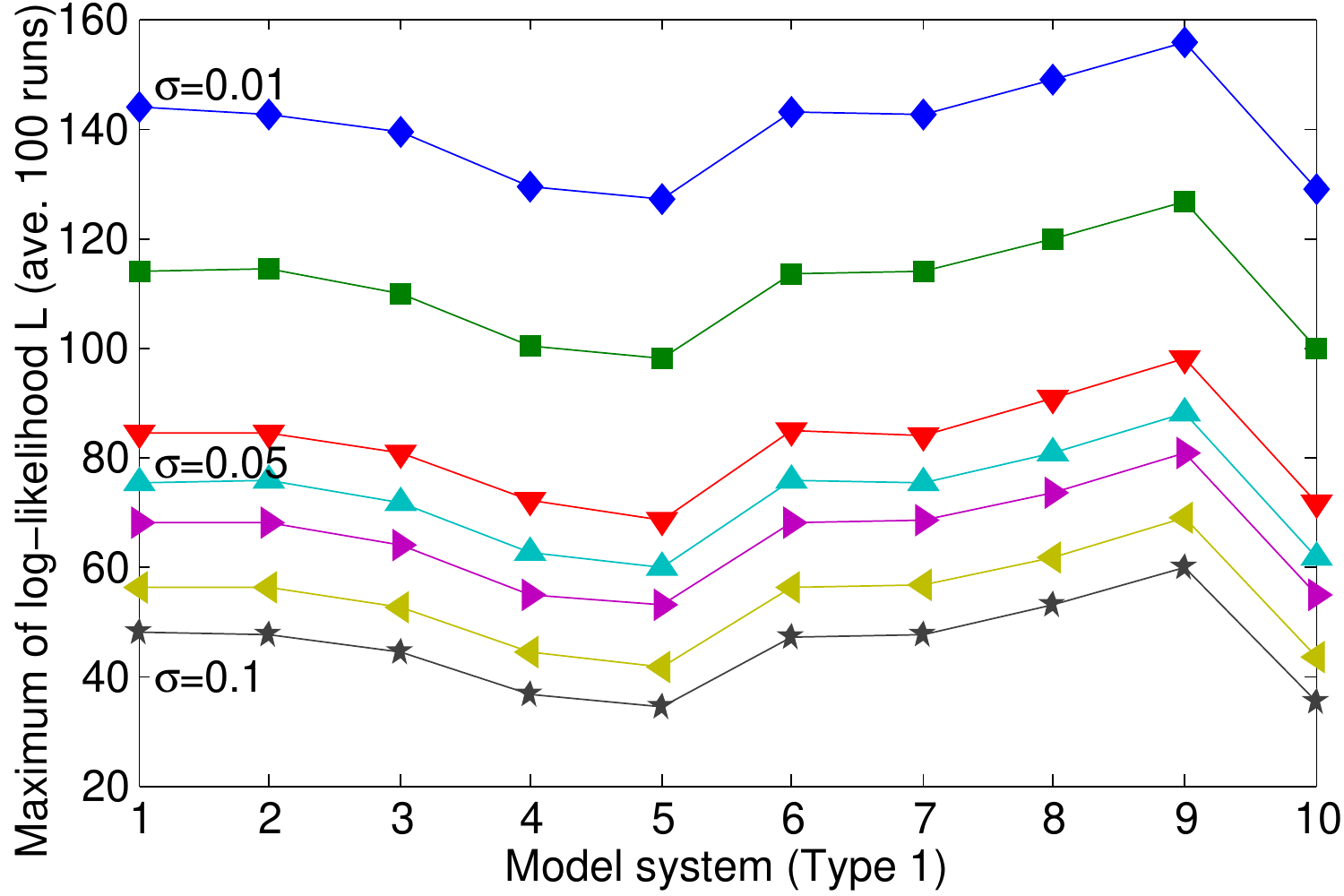}
\caption{Maximum of log-likelihood (Strategy~3) for 10 model systems
         (Table~\ref{table:models}) for different noise levels.}
\label{fig:compare1b}
\end{figure}

\subsection{Uncertainty in parameter estimates}

The error statistics are useful for comparing different strategies in
terms of both the accuracy (mean or median of error) and precision
(spread of errors) of the estimated parameters, and the graphs above
show that Strategy~3 outperforms the other strategies on both counts.
However, obtaining such statistics requires data from many simulated
experiments as well as knowledge of the actual system parameters.  In
practice, the actual values of the system parameters to be estimated
are usually unknown, as otherwise there would be no need to estimate
the parameters in the first place, so we cannot use error statistics
directly to determine the accuracy and precision of our estimates.
However, we can estimate the uncertainty of the parameter estimates,
as discussed next.

For the Fourier-based strategies we have already mentioned that the
uncertainty in the parameter estimates is mainly determined by the
frequency resolution, limited by the sampling rate based on the
Nyquist-Shannon sampling theorem, which is fixed $N_t/T$ in our case,
and the length of the sampled input signal as the Gabor limit implies
as trade-off between time- and band-limits.

For the maximum likelihood estimation we can obtain uncertainty
estimates for the parameters from the width of the peak of the
likelihood function around the maximum.  We use the following simple
strategy.  Let $(\omega,\gamma)$ be the parameters for which the
log-likelihood assumes its (global) maximum $L_{\max}$.  To estimate
the uncertainty in $\omega$ we compute the log-likelihood
$L(\omega+\delta\omega,\gamma|\vec{d})$ for values $\delta\omega$
where $L$ is significantly larger than $0$ (implemented by sampling
under the assumption that $L$ is not too far off a peaked
distribution).  Then we find the range of $\delta\omega$ for which the
actual likelihood
\begin{equation}
  \exp(L(\omega+\delta\omega,\gamma|\vec{d}) \ge \tfrac{1}{2} \exp(L_{\max})
\end{equation}
to determine the full width at half maximum (FWHM)
$\delta\omega^{\rm FWHM}$ of the likelihood peak
in the $\omega$ direction.  Assuming a roughly Gaussian peak the
uncertainty in $\omega$ is then given by
\begin{equation}
  \Delta \omega = 2 \sqrt{2 \ln(2)} \, \delta\omega^{\rm FWHM},
\end{equation}
and similarly for $\gamma$.  Fig.~\ref{fig:peak-width} shows the
resulting peaks in the likelihood function for a typical experiment
together with the FWHM estimates, showing greater uncertainty in the
$\gamma$ estimates.

Fig.~\ref{fig:compare1a} show the resulting uncertainties for
parameter estimates obtained by Strategy~3 for the ensemble
measurements.  The uncertainty in the $\omega$ and $\gamma$ estimates
increases with the noise level, as one would expect, but for some
systems the increase is steeper than for others.  In particular, the
uncertainties are greater for models 4, 5 and 10, for which $\gamma$
is large, and lowest for model system 9, which has the lowest $\gamma$
of the 10 models.  The higher uncertainties coincide with dips in the
maximum of the log-likelihood in Fig.~\ref{fig:compare1b}. Although
there is some variation in the value of the maximum log-likelihood
between different runs for the same model and error level, the
differences between the average of the maximum log-likelihood over
many runs for model systems 1 and 5 are several standard deviations,
e.g.~$\max \log L \approx 47.9 \pm 3.2$ (for model 1, $\sigma=0.1$) vs
$34.3\pm 3.3$ (model 5, $\sigma=0.1$).  This is consistent with the
peak of the (log-)likelihood being lower and broader for model 5,
resulting in higher uncertainty, and narrower and higher for model 1,
resulting in less uncertainty.  Fig.~\ref{fig:compare2a} shows that
the uncertainties for parameter estimates behave the same ways for
single shot measurements as a function of the projection noise level
$1/\sqrt{N_e}$.

This suggests that given the same amount of data the uncertainty of
our estimates increases slightly with larger dephasing rate.  A
probable explanation for this is that the signal decays faster for
higher dephasing and thus the signal-to-noise ratio of the later time
samples is reduced.  For higher dephasing rates the results could
likely be improved by adding more samples for shorter times or
introducing weights and reducing the latter for measurements obtained
for longer times.

\begin{figure}
\includegraphics[width=0.49\textwidth]{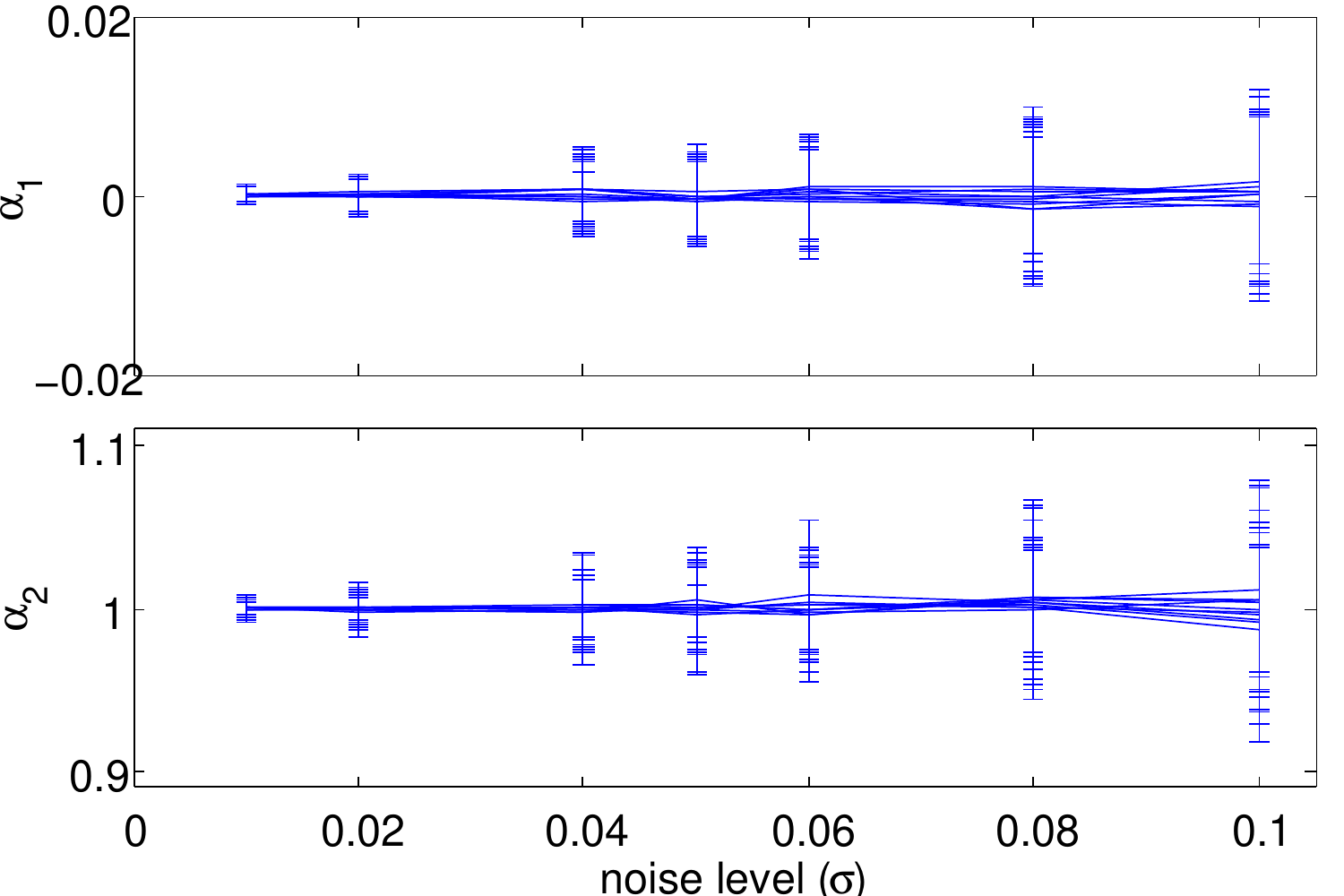}
\caption{Estimates for parameters $\alpha_1$ and $\alpha_2$ including
  uncertainty as a function of the noise level $\sigma$ for 10 model
  systems.}
\label{fig:compare1d}
\end{figure}

\begin{figure}
\includegraphics[width=0.49\textwidth]{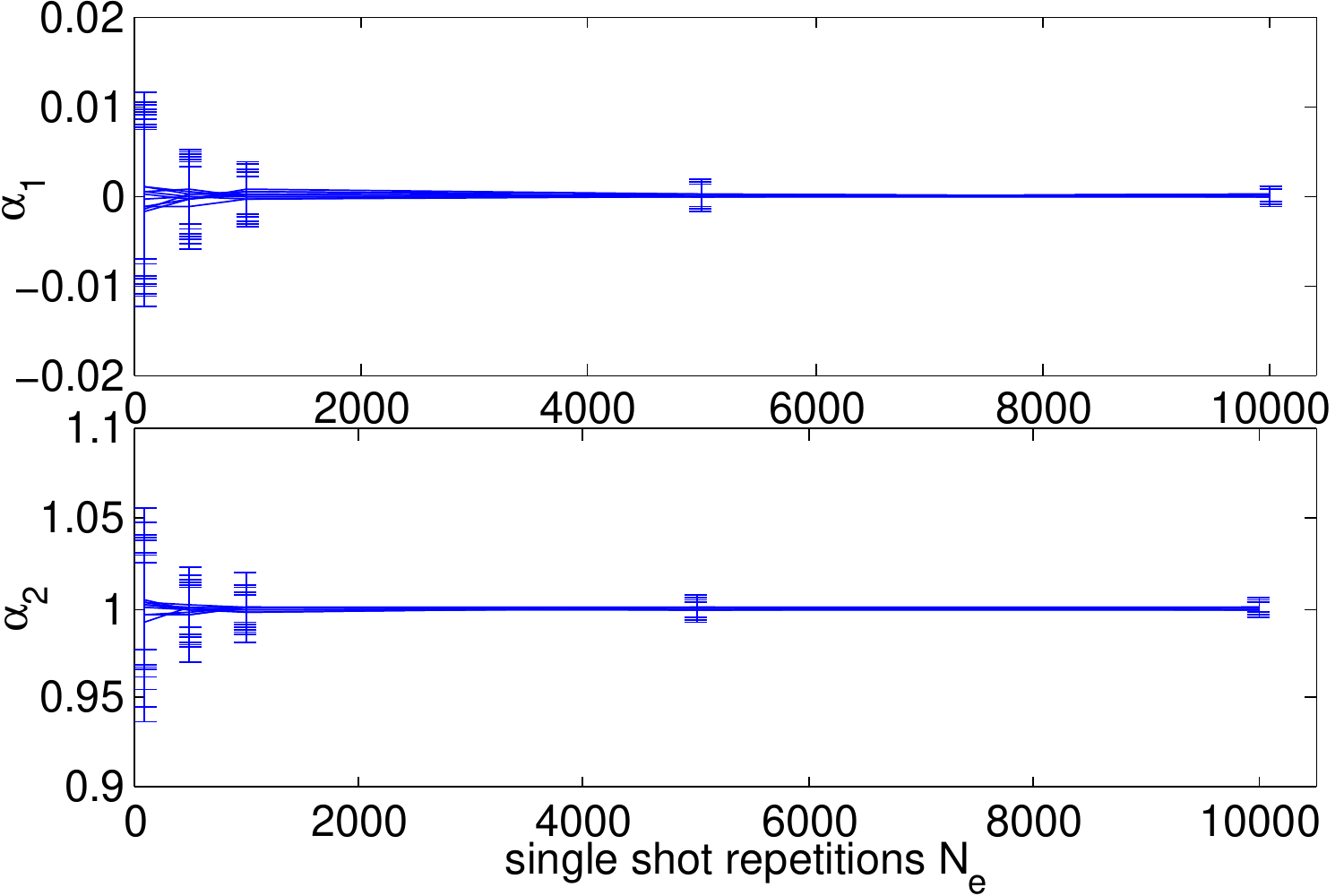} \\
\caption{Estimates for parameters $\alpha_1$ and $\alpha_2$ as a
  function of the number of single shot repetitions $N_e$ for 10 model
  systems (Type 1, averaged over 100 runs each).}
\label{fig:compare2d}
\end{figure}

\subsection{Estimating initialisation and measurement parameters}

According to (\ref{eq:im}) Strategy~3 also provides information about
the initialization and measurement procedure via estimates for the
parameters $\alpha_1$ and $\alpha_2$. For this model we obtain
\begin{equation*}
\alpha_1 \pm \alpha_2 = \cos\theta_I \cos\theta_M \pm \sin\theta_I \sin\theta_M
                      = \cos(\theta_I\mp\theta_M)
\end{equation*}
and thus
\begin{subequations}
\begin{align}
 \theta_I &= \tfrac{1}{2}[\arccos(\alpha_1 - \alpha_2) + \arccos(\alpha_1+\alpha_2)], \\
 \theta_M &= \tfrac{1}{2}[\arccos(\alpha_1 - \alpha_2) - \arccos(\alpha_1+\alpha_2)].
\end{align}
\end{subequations}

Fig.~\ref{fig:compare1d} shows the estimates for the parameters
$\alpha_1$ and $\alpha_2$ with error bars indicating uncertainty for
the ensemble measurements.  From the plot it is evident that
$\alpha_1\to0$ and $\alpha_2\to1$ for $\sigma\to 0$, which suggests
$\theta_I=\theta_M=\tfrac{\pi}{2}$, which agrees with the values of
the initialization and measurement angles used in the simulated
experiments.  Fig.~\ref{fig:compare2d} shows that the same is true in
the case of projection noise for single shot measurements.  The
associated estimates for the parameters $\alpha_1$ and $\alpha_2$ in
converge to $\alpha_1\to0$ and $\alpha_2\to1$ for $N_e\to \infty$,
which suggests $\theta_I=\theta_M=\tfrac{\pi}{2}$, which also agrees
with the values of the initialization and measurement angles used in
the simulated experiments.  Similar behaviour is observed for other
choice of the initialization and measurement angles.

\subsection{Fisher Information and Cramer Rao Bound}

The Fisher information matrix $I=(I_{ij})$ is defined by
\begin{equation}
%\begin{split}
   I_{ij} =  E\left[ \frac{\partial L}{\partial\theta_i} \frac{\partial L}{\partial\theta_j} \right]
           =  \int \frac{\partial L}{\partial\theta_i} \, \frac{\partial L}{\partial\theta_j} f(x|\theta) dx
          = - E \left[ \frac{\partial^2L}{\partial \theta_i \partial\theta_j} \right]
%\end{split}
\end{equation}
where $L(x,\theta)$ is the log-likelihood of the
measurement outcome $x$ given $\theta$ and $E$ the expectation
w.r.t. $x$.  If the estimator $T$ for the parameters $\theta$ is
unbiased, i.e. the mean square error of $T$ is
\begin{equation}
  \mbox{\rm MSE}(T) = \mbox{\rm Bias}(T)^2 + \Var(T) = \Var(T)
\end{equation}
where $\Var(T)$ is the covariance matrix of the estimator,
then the matrix $C=\Var(T)-I^{-1}$ must be positive semi-definite and $\norm{C}$
gives an estimate of how close we are to the Cramer-Rao limit.

\begin{figure}
\includegraphics[width=0.49\textwidth]{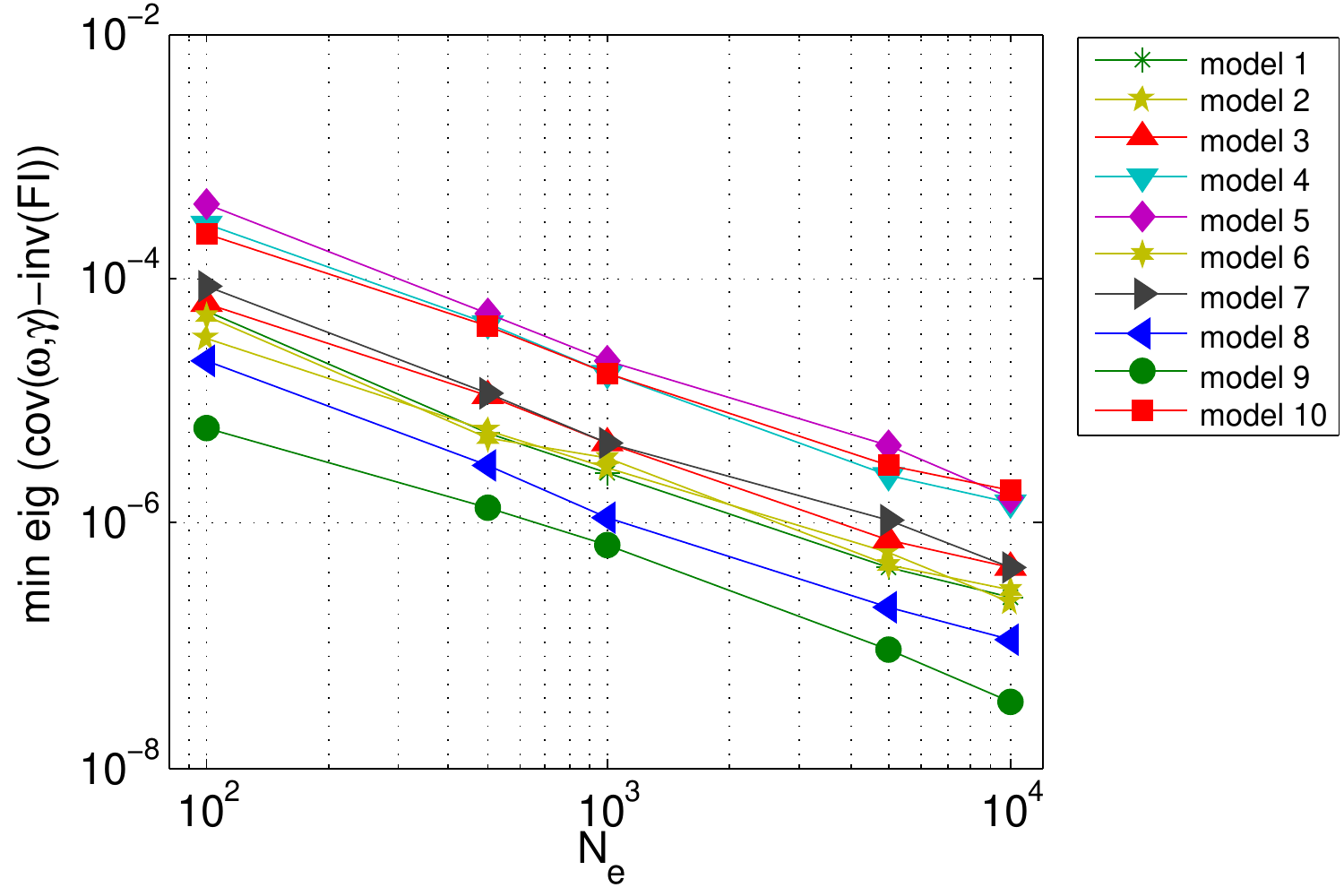}
\caption{Plot of the minimum eigenvalue of the covariance matrix of
  the estimator minus the inverse Fisher information for various
  models as a function of $N_e$.}
\label{fig:fisher}
\end{figure}

Applied to our case, $\theta = (\omega,\gamma)$ and
\begin{equation*}
  L(\vec{x}|\theta) = -N \log (\sqrt{2\pi} \sigma) - \frac{1}{2\sigma^2}\sum_{n=1}^N|p(\theta,t_n) - x_n|^2
\end{equation*}
with $p(\theta,t) = e^{-\theta_2 t} \cos(\theta_1 t)$, we get
\begin{subequations}
\begin{align}
 \frac{\partial L}{\partial \theta_1}
 & = -\frac{1}{\sigma^2} \sum_{n=1}^N [p(\theta,t_n)-x_n] \frac{\partial p(\theta,t_n)}{\partial \theta_1}\\
\frac{\partial L}{\partial \theta_2}
 & = -\frac{1}{\sigma^2} \sum_{n=1}^N [p(\theta,t_n)-x_n] \frac{\partial p(\theta,t_n)}{\partial \theta_2}\\
\end{align}
\end{subequations}
and
\begin{subequations}
\begin{align}
  \frac{\partial p(\theta,t_n)}{\partial \theta_1}
  &= - t_n e^{-\theta_2 t_n} \sin(\theta_1 t_n) = : \alpha_n\\
  \frac{\partial p(\theta,t_n)}{\partial \theta_2}
  &= - t_n e^{-\theta_2 t_n} \cos(\theta_1 t_n) = : \beta_n.\\
\end{align}
\end{subequations}
Setting $p_n = p(\theta,t_n)$ we have
\begin{align*}
  \frac{\partial L}{\partial \theta_1} \frac{\partial L}{\partial \theta_2}
 &= \frac{1}{\sigma^4} \left( \sum_{n=1}^N \alpha_n p_n -\alpha_n x_n \right)
                       \left( \sum_{n=1}^N \beta_n p_n - \beta_n x_n \right)\\
 &= \sigma^{-4} \left( AB - \sum_{n=1}^N c_n x_n  + \sum_{m,n=1}^{N} \alpha_m \beta_n x_m x_n \right)
\end{align*}
with $A = \sum_n \alpha_n p_n$ and $B = \sum_n \beta_n p_n$,
$c_n=\alpha_n B + \beta_n A$.  Similarly for the other partial
derivatives.  Noting
\begin{equation}
   \frac{1}{\sqrt{2\pi} \sigma} \int_{-\infty}^\infty x_n \exp \left[ \frac{-|p_n-x_n|^2|}{2\sigma^2} \right]
    d x_n = p_n
\end{equation}
and assuming the estimator is unbiased, we finally obtain the
entries of the Fisher information matrix
\begin{equation}
\begin{split}
 I_{11} &= \sigma^{-4}  \left(A^2 - 2A\sum_n \alpha_n p_n + \sum_{m,n} \alpha_m \alpha_n p_m p_n \right)\\
 I_{12} &= \sigma^{-4}  \left(AB  - \sum_n c_n p_n + \sum_{m,n} \alpha_m \beta_n p_m p_n \right)\\
 I_{22} &= \sigma^{-4}  \left(B^2 - 2B\sum_n \beta_n p_n + \sum_{m,n} \beta_m \beta_n p_m p_n \right).
\end{split}
\end{equation}
While our simulations suggest that the estimators based on Strategies~1
and~2 are not unbiased, Strategy~3 appears to be unbiased.
Fig.~\ref{fig:fisher}, showing the smallest eigenvalue of the matrix
$C$ for our various test systems subject to projection noise,
suggests that we indeed approach the Cramer-Rao bound for $N_e\to\infty$
and $\sigma=N_e^{-1/2}$.

\section{Adaptive Estimation Strategies}

\begin{figure*}
\includegraphics[width=0.32\textwidth]{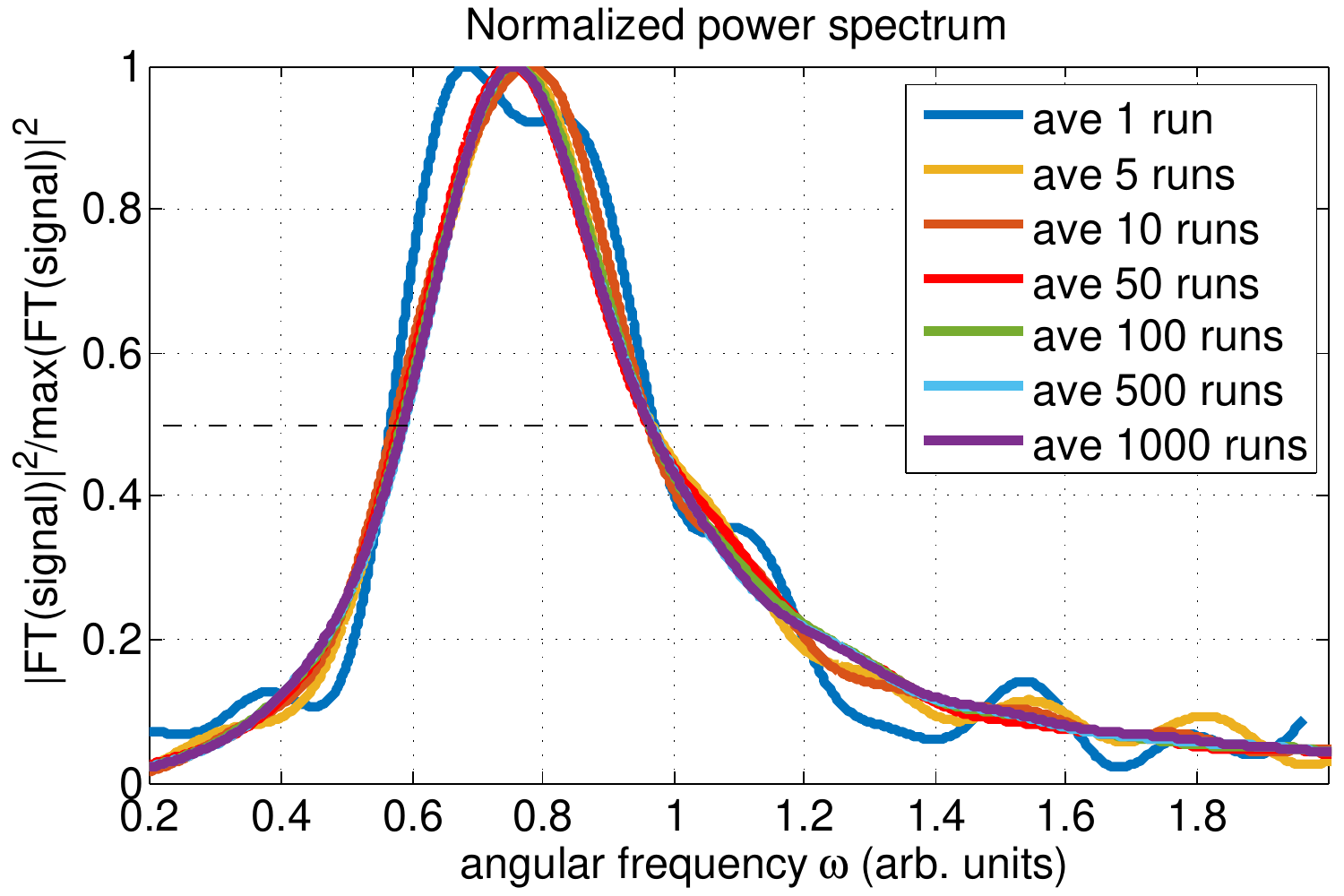}
\includegraphics[width=0.32\textwidth]{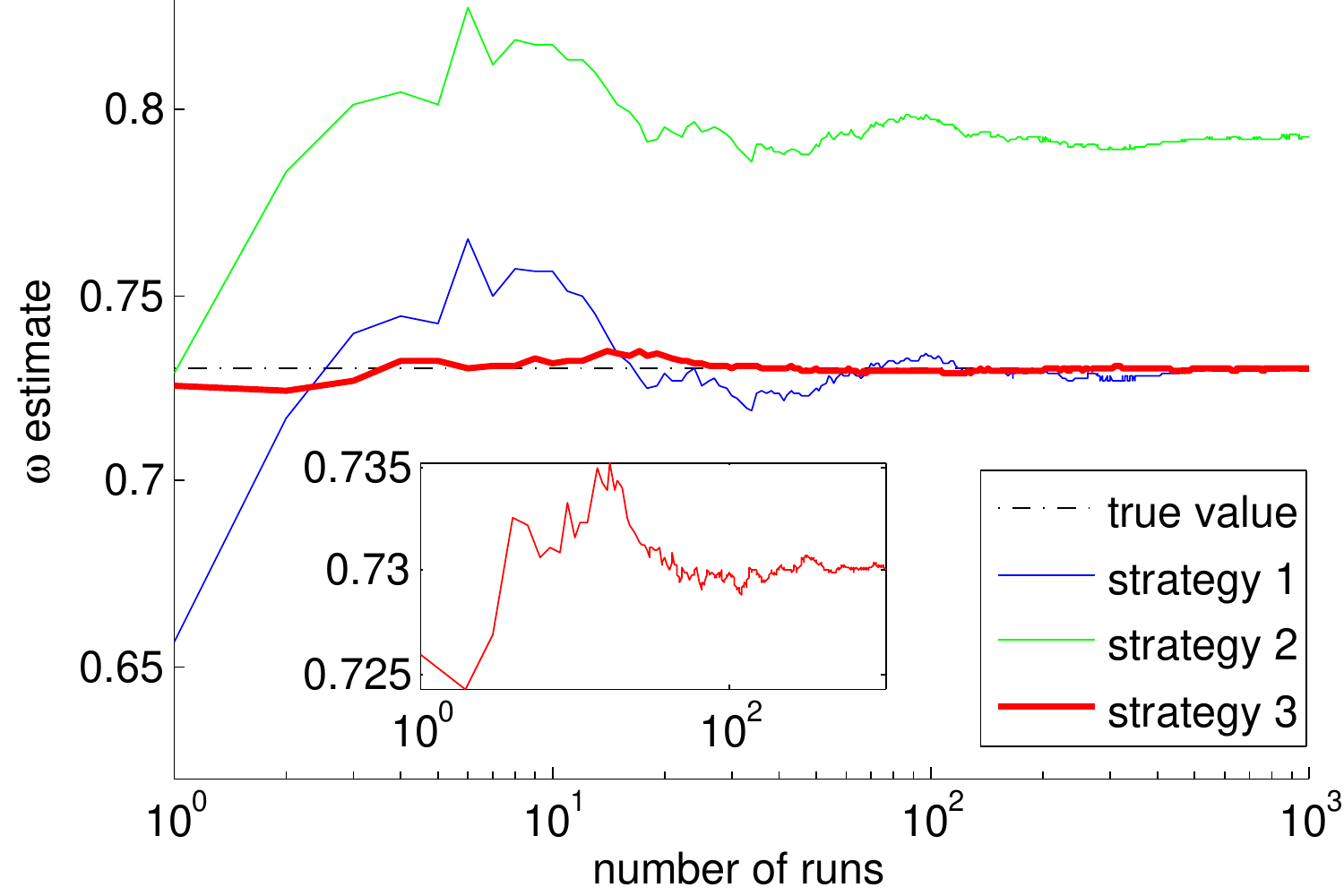} \hfill
\includegraphics[width=0.32\textwidth]{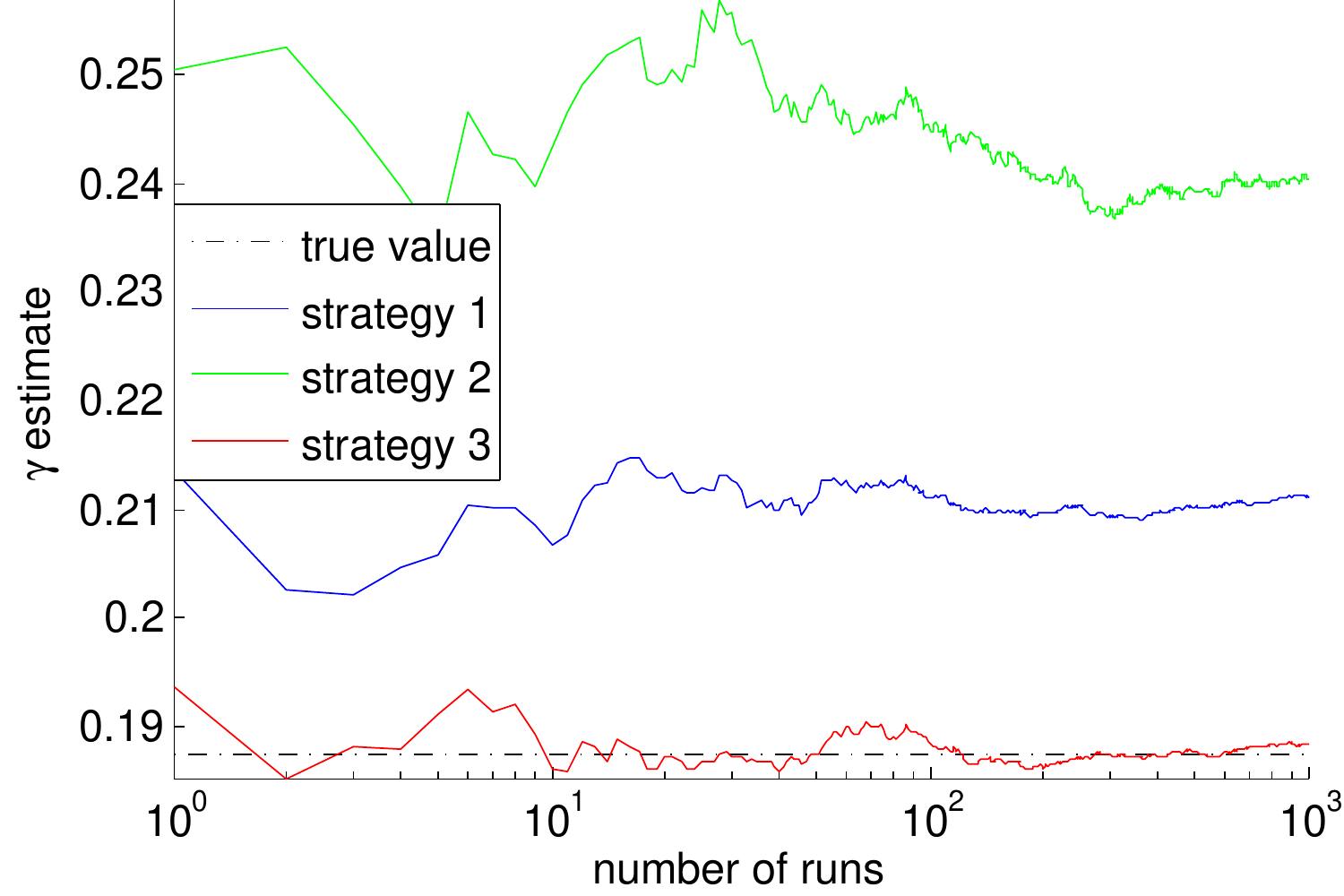} \hfill
\caption{Iterative refinement by averaging of signal traces: power
  spectra (left), $\omega$ estimates (center) and $\gamma$ estimates
  (right). }
\label{fig:refine1}
\end{figure*}

\begin{figure*}
\includegraphics[width=0.49\textwidth]{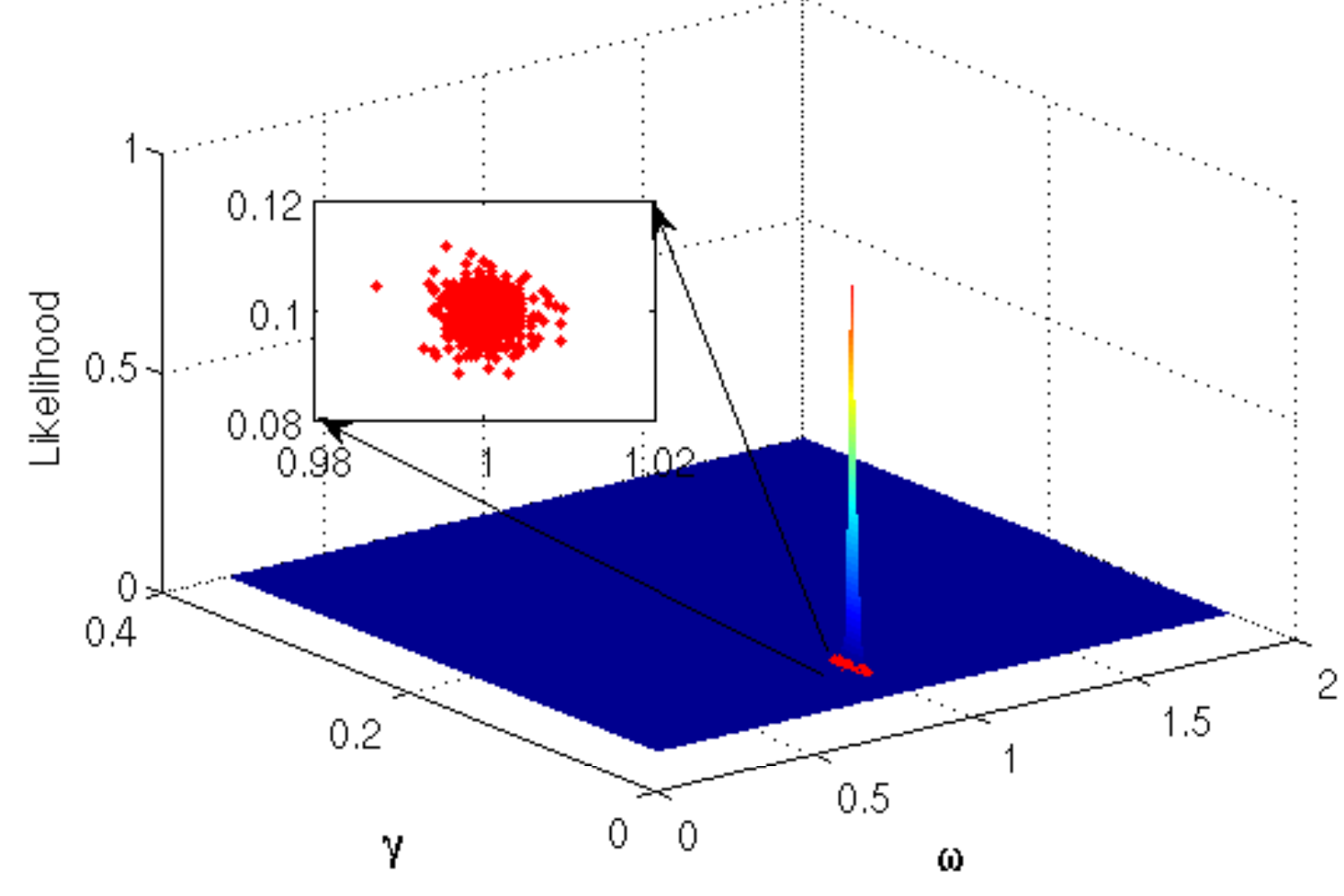} \hfill
\includegraphics[width=0.49\textwidth]{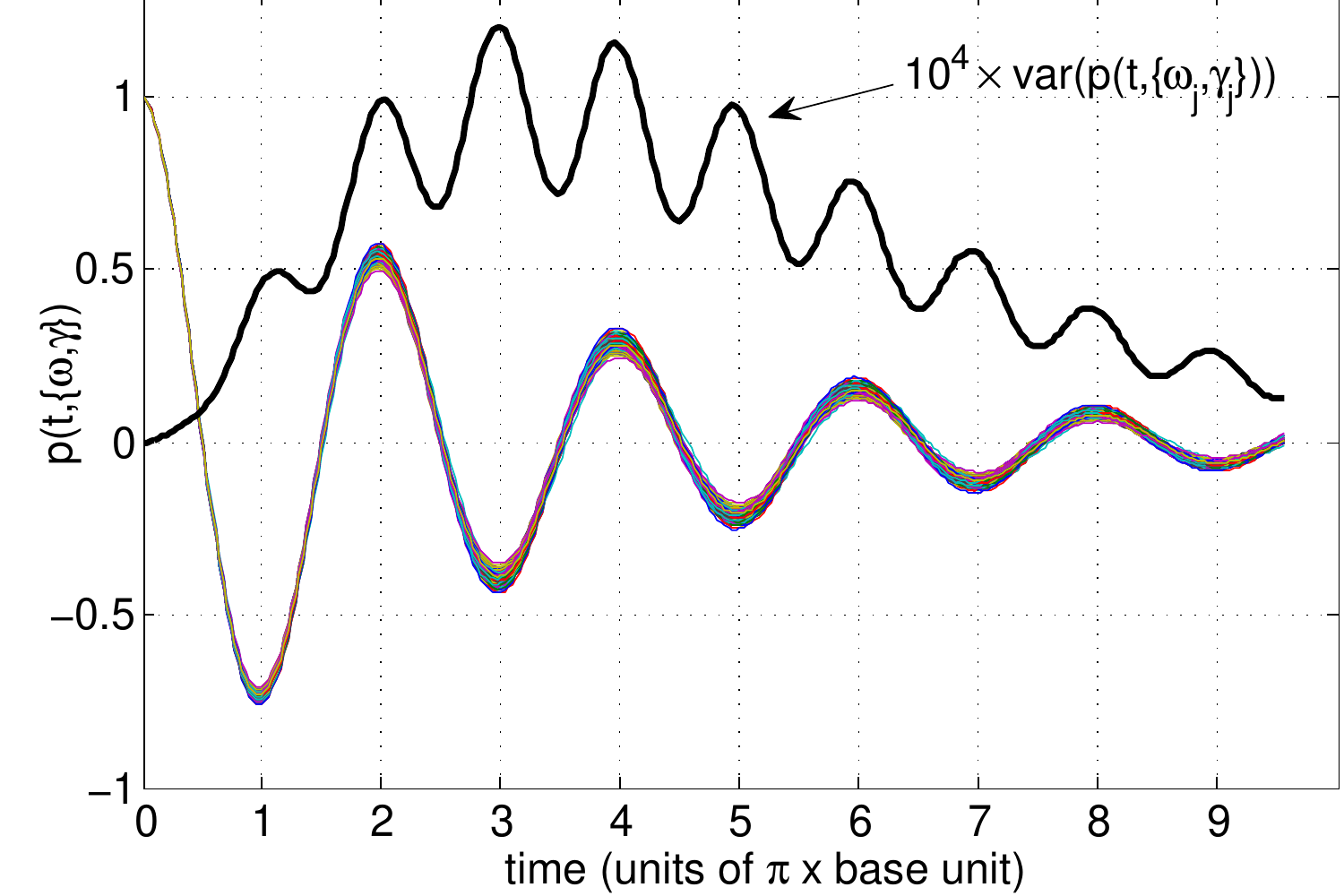}
\caption{Prior likelihood after 25 time samples ($t_n=1.2n$) for model 1 (left)
with $\{\omega_j,\gamma_j\}$ samples (red dots) and corresponding predicted
measurement traces $p_j(t) = p(t,\{\omega_j,\gamma_j\})$ and variance of
$p_j(t)$ as function of $t$ (right).}
\label{fig:refine2}
\end{figure*}

We may find that the accuracy or precision of the parameters obtained
from an initial data set is not sufficient and we would like to
improve it by acquiring additional data.  Adaptive refinement
strategies depend on the experimental set-up and system and a detailed
analysis of specific strategies is beyond the scope of this paper.
However, we shall briefly discuss general approaches for iterative
refinement for the Fourier and Bayesian estimation approaches and
compare these for a few examples.

In some settings an entire measurement trace is obtained in a single
experimental run and we are only able to sample the signal at regular
time intervals restricted by the experimental equipment available.  In
this case the only options available to us are extending the signal
length (keeping sampling density or number of sample points constant)
or repeating the experiment.  If Fourier-based estimation strategies
are used, the only way to really improve the resolution of the Fourier
spectrum, and thus the accuracy and precision of our estimates, is by
increasing the signal length. However, for a decaying signal the
signal-to-noise ratio progressively deteriorates until the signal
vanishes, limiting the accuracy and precision that are
attainable. This is illustrated in Fig.~\ref{fig:refine1}(left), which
shows the (normalized) power spectrum for 1 to 1000 repetitions of the
experiment for model parameters 4, assuming each individual
measurement trace is subject to Gaussian noise at $\sigma=0.1$ and the
signals are averaged. For a single run of the experiment with this
level of noise, the peak is distorted but the power spectrum quickly
converges.  The corresponding estimates for $\omega$ and $\gamma$
(Fig.~\ref{fig:refine1}, center and right) also converge but not to
the true value.  For Strategy~2 the $\omega$ and $\gamma$ estimates
are inaccurate.  The optimization step in Strategy~1 appears to
improve the accuracy of the $\omega$ estimates but the $\gamma$
estimates are still inaccurate.  Strategy~3 does not suffer from these
limitations and averaging multiple short traces should increase the
accuracy of our estimates.  Indeed the figure shows that this appears
to be the case: both the $\omega$ and $\gamma$ estimates converge to
the true values.

This shows that Strategy~3 allows adaptive refinement even if all we
are able to do is to repeat the experiment multiple times and average
the measurement traces.  However, in some situations we have more
freedom.  For Rabi spectroscopy, for example, each data point,
corresponding to a measurement at a particular time $t_n$, may be
obtained in a separate experiment, and we may be free to choose the
measurement times $t_n$ flexibly.  In this case, having obtained $N_t$
measurements we can try to choose the next measurement time
$t_{N_t+1}$ such that it optimizes the amount of information we gain
from the experiment.  We could ask, for example, considering all
possible outcomes of a measurement at time $t$ and their probability
based on our current knowledge, at what time should we measure next to
achieve the largest reduction in the uncertainty of our estimates.
However, this would require calculating the uncertainty of the
parameters (e.g., by estimating the width of the likelihood peaks) for
all possible measurement times and outcomes.  Given the continuum of
measurement outcomes and measurement times, this is generally too
expensive to calculate.

We therefore consider a simpler heuristic.  We generate a number of
guesses $\{(\omega_j,\gamma_j), j=1,\ldots, J\}$ for the parameters
based on the current likelihood distribution for the parameters.  We
then calculate the measurement signal $p(t,\{\omega_j,\gamma_j\})$ for
a set of discrete times and select the next measurement time where the
variance of the predicted measurement results is greatest.  The idea
behind this strategy is that a larger spread in the predicted results
indicates greater uncertainty, and a measurement at such a time should
result in a greater reduction of the uncertainty.  We illustrate this
strategy in Fig.~\ref{fig:refine2}.  The variance of the predicted
traces $p_j(t) = p(t,\{\omega_j,\gamma_j\})$ exhibits oscillations at
about twice the frequency of the signal, being largest around the
minima and maxima of the oscillatory signal but due to the damping of
the signal there is an overall envelope and a global maximum around
$3$ in units of $\pi\bar{\omega}^{-1}$.  To avoid repeated sampling at
the same time it is desirable to introduce a degree of randomness,
e.g., by selecting the next measurement time based on the maximum of
the variance of $p_j(t_s)$ sampled over a discrete set of times $t_s$,
such as a non-uniform low-discrepancy sampling of the time interval
$[0,T]$.  Furthermore, in practice it may be rather inefficient to
recalculate the variance of the traces after a single measurement.
Instead, it we shall acquire an initial set of $N_0$ data points and
then select the next $N_1$ measurement times to coincide with peaks in
the variance of the traces where we allow $N_1$ to vary depending on
the number of peaks.  In Fig.~\ref{fig:refine2}, for example, there
are eight local peaks and we would choose the next eight measurement
times to coincide with these maxima and iterate the process.

\begin{figure}
\includegraphics[width=0.49\textwidth]{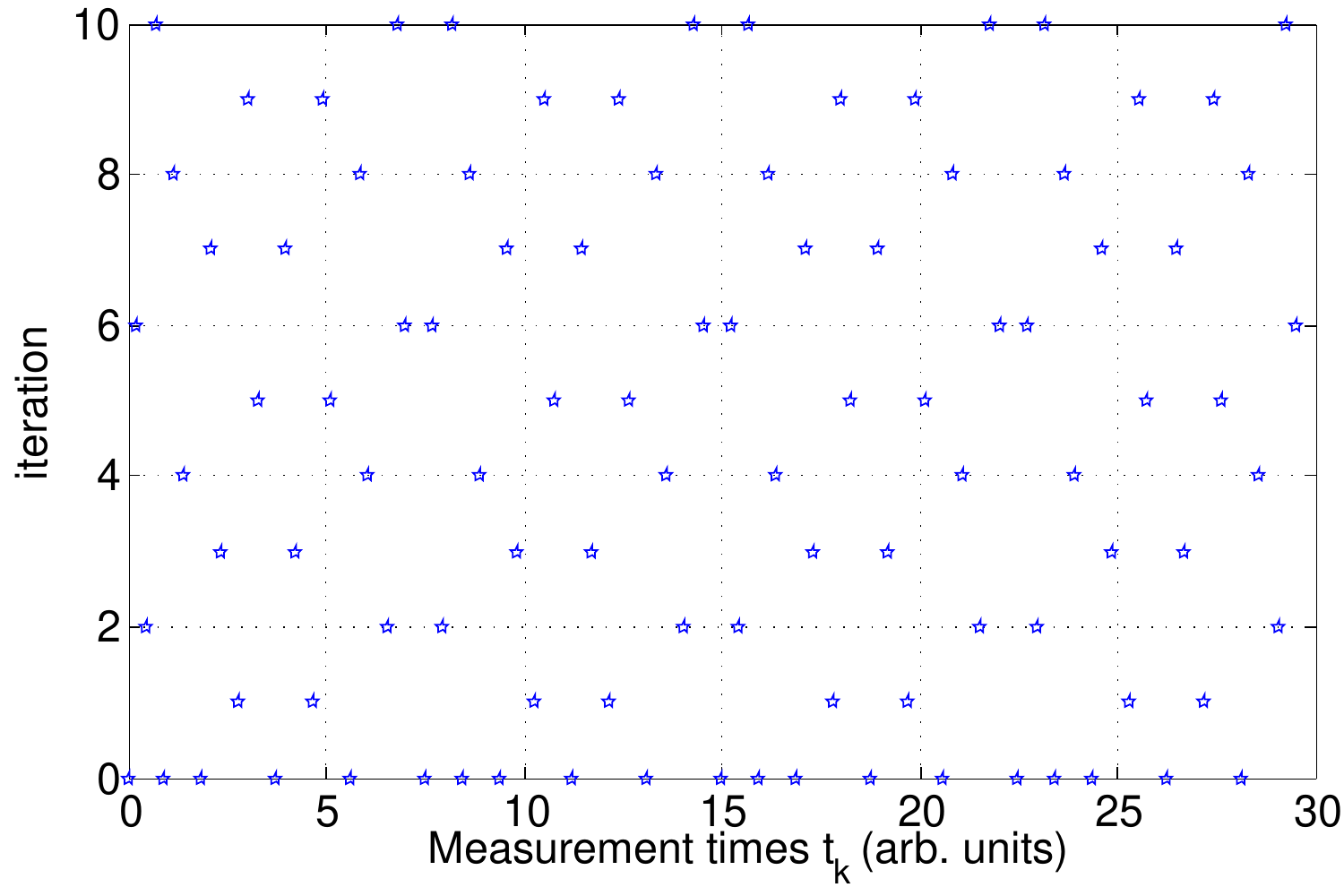}
\caption{Selection of measurement times for interative low-discrepancy
  sampling.  The new measurement times in each iteration as chosen
  such as to fill in the largest existing gaps.}
\label{fig:ld-iter}
\end{figure}

\begin{figure}
\includegraphics[width=0.49\textwidth]{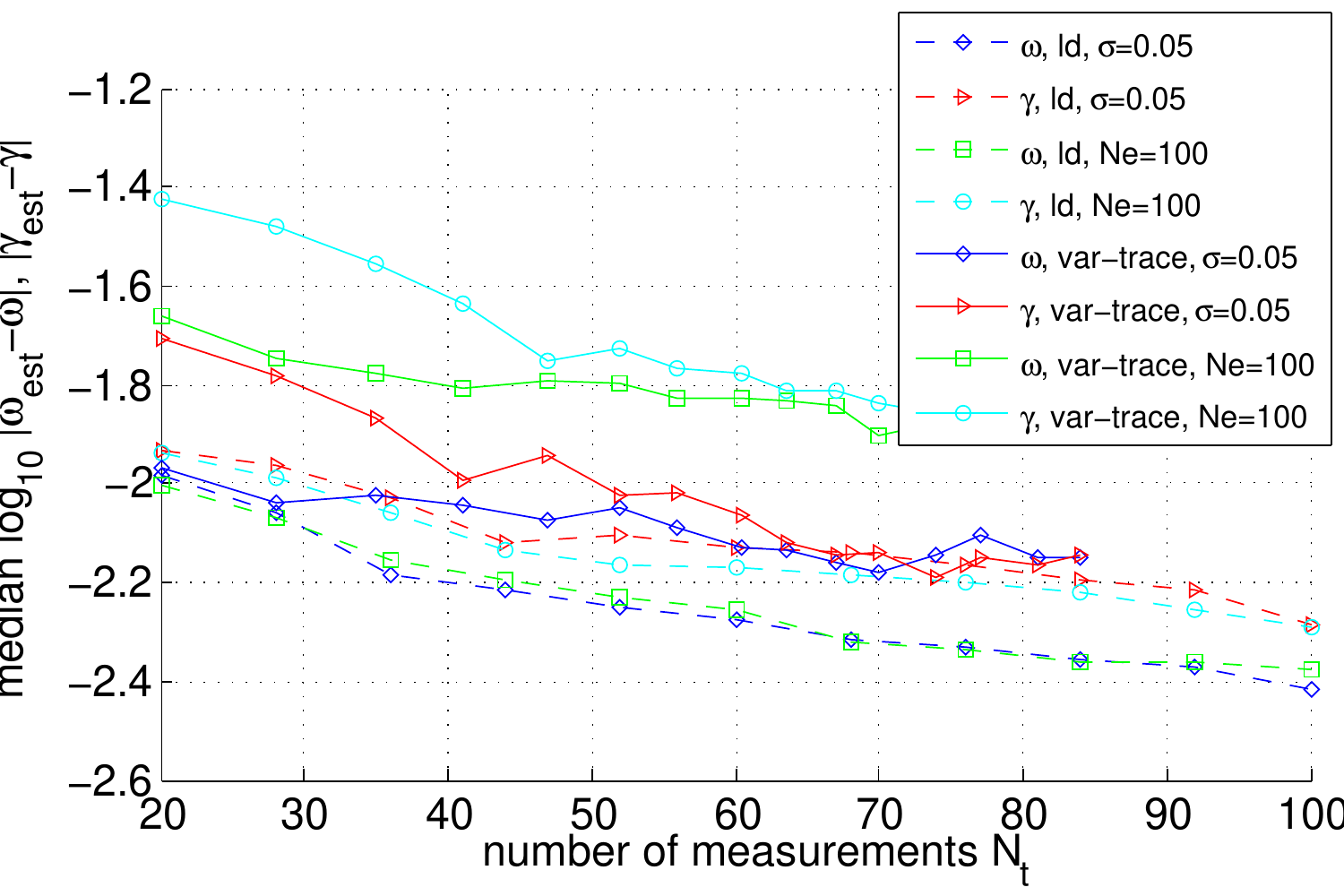}
\caption{Median error of $\omega$ and $\gamma$ parameter estimates for
  iterative ld-sampling and adaptive sampling based on trace variance
  for model system 4 with measurements subject to 5\% Gaussian noise
  and projection noise $\sigma=N_e^{-1/2}$, respectively.}
\label{fig:ld-refine}
\end{figure}

An even simpler way of iterative refinement is via low-discrepancy
(ld) time sampling, a generalization of uniform sampling that lends
itself to easy iterative refinement. The basic idea of ld sequences
is to ensure the largest gap between samples is asymptotically optimal,
while there is little uniformity in the sampling points to avoid
aliasing effects (see blue noise criterion). In this case the initial
measurement times are chosen to be the first $N_0$ elements in a
low-discrepancy quasi-random sequence such as the Hammersley
sequence~\cite{Hammersley}, and in each subsequent iteration the next
$N_i$ elements of the sequence are used.  The number of initial
measurements $N_0$ and subsequent measurements per iteration $N_i$ are
completely flexible, the elements of the sequence can be scaled to
uniformly cover any desired time interval, and we can perform as many
iterations as desired.  Fig.~\ref{fig:ld-iter} shows the measurement
times as a function of the iteration as determined by the Hammersley
sequence with $N_0=20$ and $N_i=8$ for 10 iterations and total
sampling times $T=30$, showing that uniform coverage of the sampling
interval is maintained.  For a fixed number of measurements $N_t=100$
we verified that there was no significant difference in the errors and
uncertainties of the parameter estimates between low-discrepancy and
uniform sampling for the cases considered above.  Furthermore,
iterative refinement based on ld-sampling performed very well.
Fig.~\ref{fig:ld-refine} for model system 4 with measurements subject
to 5\% Gaussian noise shows that simple iterative ld-sampling actually
outperforms the adaptive refinement strategy based on the
trace-variance described above.  While this may not be universally the
case, and may be due to the variations in the trace variance being
relatively small in our example, it shows that simple strategies such
as iterative ld-sampling can be highly effectively.

%F Still the question of how this comapres with uniform sampling... (which is also ld, just has a blue noise problem, so issues with aliasing artefacts)
%F Can we say this suggests that even, but aliasing artefact minimal, sampling is actually really what is needed instead of high repetition or VoI, etc?

\section{Generalization to Other Models}

\begin{figure*}
{\includegraphics[width=0.49\textwidth]{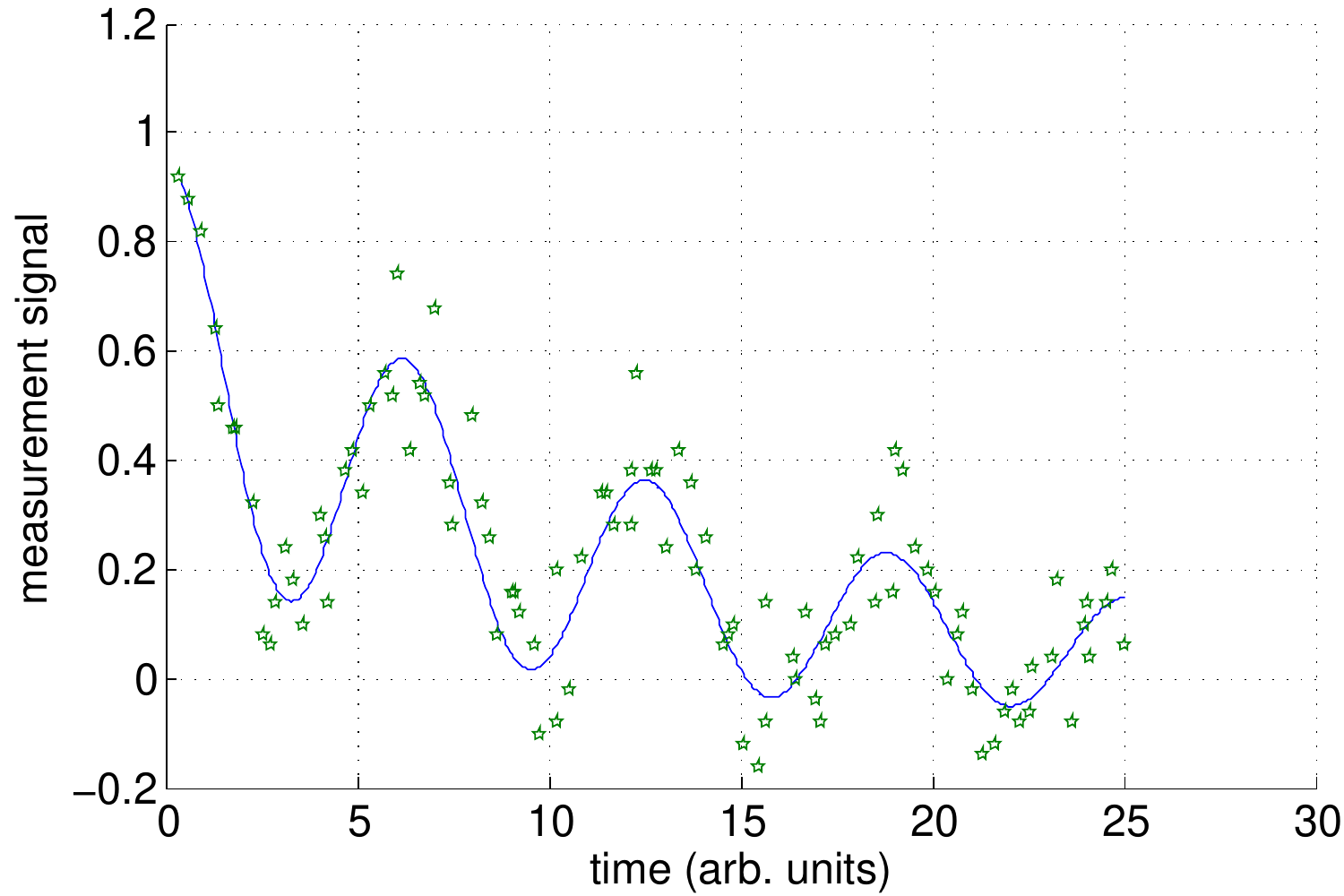}} \hfill
{\includegraphics[width=0.49\textwidth]{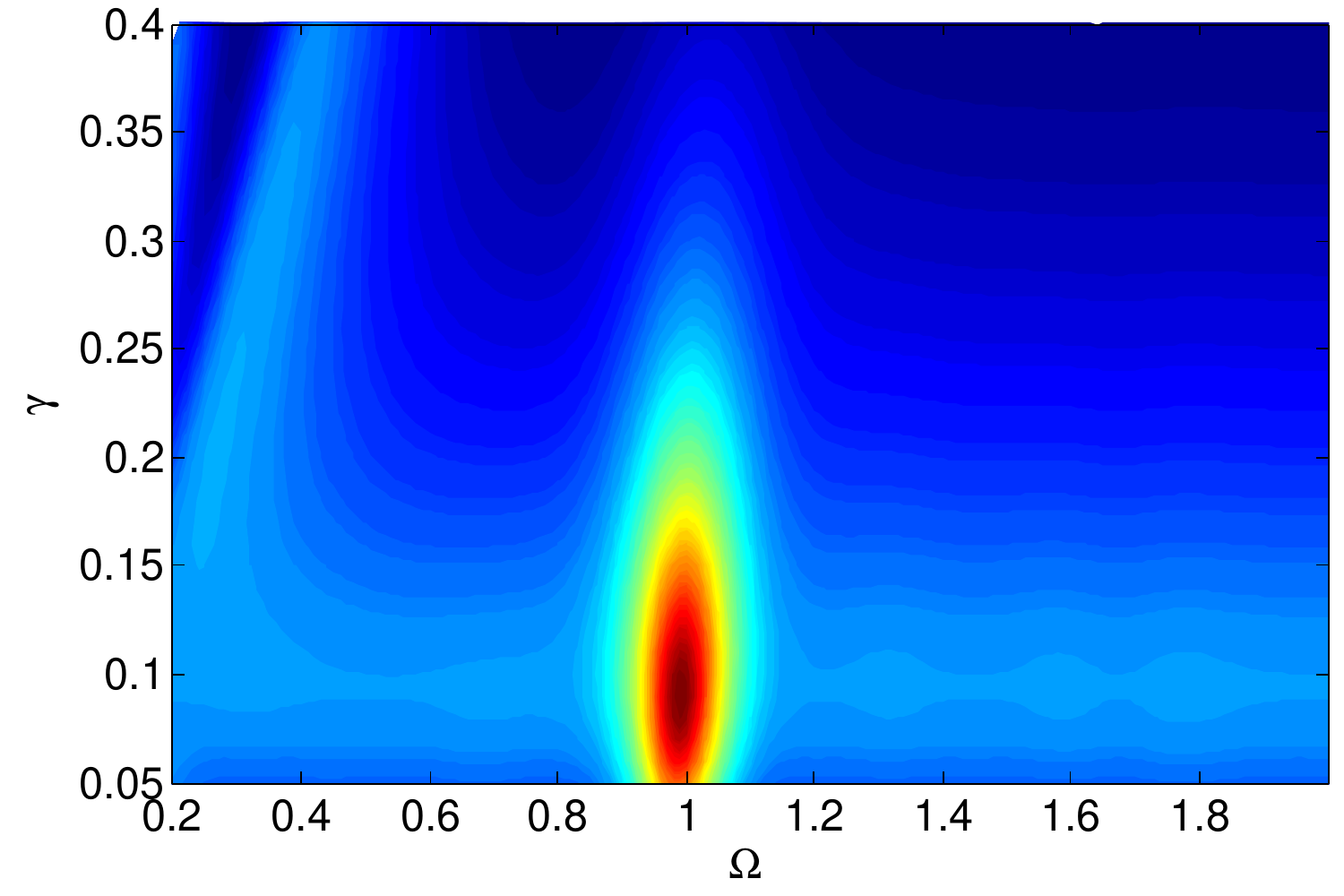}}
\caption{Ideal signal (blue) and sparsely sampled noisy data (green
  $\ast$, $N_t=100$, $t\in[0,25]$, $N_e=100$ single shot experiments per
  data point) for a system described by Eq.~(\ref{eq:meas2}) with
  $\omega=1$, $\gamma=0.1$ (left) and corresponding log-likelihood
  (right).}
\label{fig:type2}
\end{figure*}

\begin{figure*}
\includegraphics[width=0.49\textwidth]{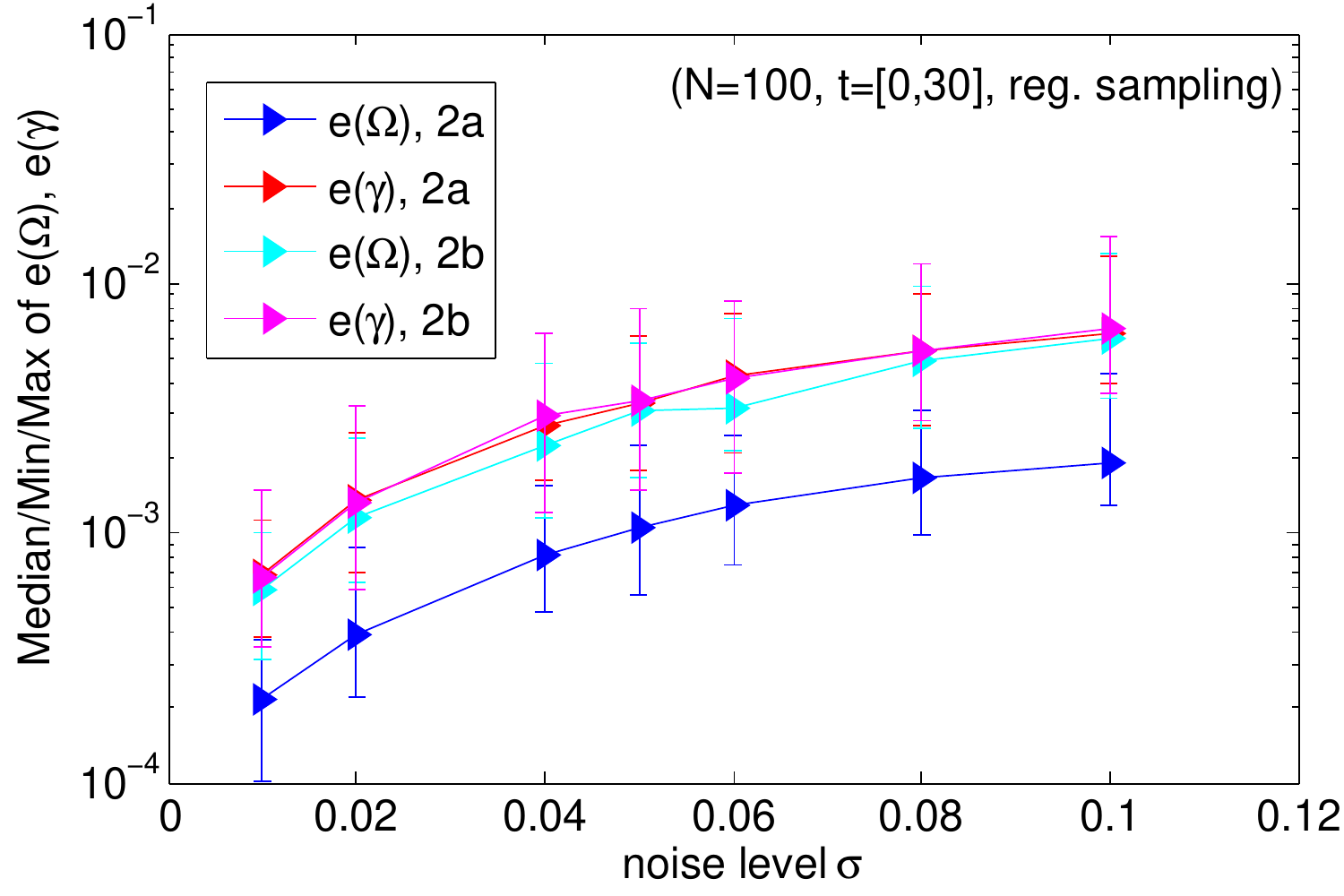} \hfill
\includegraphics[width=0.49\textwidth]{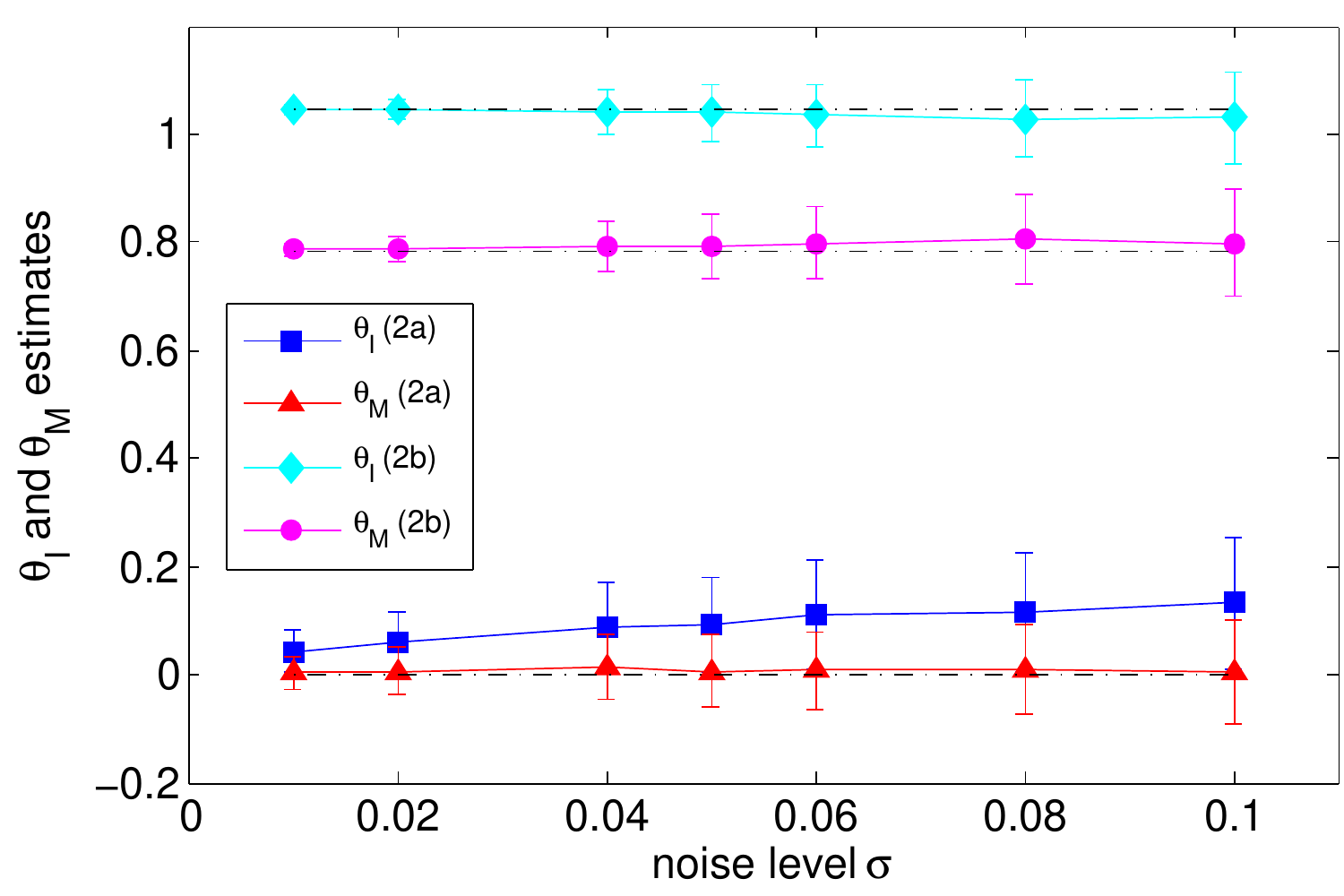}
\caption{Minimum, maximum and median of relative error (averaged over
  100 runs for each system and noise level) of $\omega$ and $\gamma$
  estimates as a function of noise level $\sigma$ (left) and estimates
  for the initial state and measurement angles $\theta_I$ and
  $\theta_M$ (right) for $10$ model systems of type (\ref{eq:meas2})
  with model parameters given in (Table~\ref{table:models}) for two
  experimental conditions: $\theta_I=\theta_M=0$ (2a, maximum
  visibility) and $\theta_I=\tfrac{\pi}{3}$, $\theta_M=\tfrac{\pi}{4}$
  (2b).}
\label{fig:type2-error}
\end{figure*}

So far we have considered a particular model of a dephasing two-level
system with dephasing acting in the Hamiltonian basis.  However, if
control fields are applied, as in a Rabi oscillation experiment for
example, then the effective Hamiltonian and the dephasing basis may
not coincide. For example, for two-level atoms in a cavity driven
resonantly by a laser, the effective Hamiltonian with regard to a
suitable rotating frame is $H=\Omega\sx$, where $\Omega$ is the Rabi
frequency of the driving field.  Assuming the driving field does not
alter the dephasing processes, so that we still have
$V=\sqrt{\tfrac{\gamma}{2}}\sz$, the resulting measurement trace is
given by~\cite{Gong2014}:
\begin{equation}
\label{eq:meas2}
  p(t) = e^{-\gamma t}\sin\theta_I\sin\theta_M +\Phi^x_3(t)\cos\theta_I\cos\theta_M
\end{equation}
where
\begin{align}
 \Phi^x_3(t) &= e^{-\tfrac{\gamma}{2}{t}} \left[\cos(\omega t)
                    +\frac{\gamma}{2 \omega}\sin(\omega t) \right], \\
  \omega     &= \sqrt{\Omega^{2}-\tfrac{\gamma^{2}}{4}} \label{omegahat}.
\end{align}
If $\Omega^2<\gamma^2/4$ then $\omega$ is purely imaginary and the
sine and cosine terms above turn into their respective hyperbolic sine
and cosine equivalents.  If $\Omega^2=\gamma^2/4$, the expression
$\omega^{-1}\sin(\omega t)$ must be analytically continued.

Due to the more complex nature of the signal, the Fourier estimation
strategies are not directly applicable.  However, we can very easily
adapt Strategy~3.  All that is required is a change in the basis
functions, setting $g_1(t)=e^{-\gamma t}$ and $g_2(t) = \Phi^x_3(t)$.

Fig.~\ref{fig:type2} shows the log-likelihood functions for a very
sparsely sampled signal with significant projection noise for a system
of type (\ref{eq:meas2}) for a simulated experiment performed with
$\theta_M=\tfrac{\pi}{4}$ and $\theta_I=\tfrac{\pi}{3}$.  The signal
is a damped oscillation, though not a simple damped sinusoid.
Strategy~3 easily succeeds in identifying the model
parameters and the log-likelihood function has a clearly defined peak.
In fact, we are showing the log-likelihood here as the actual likelihood
function is so sharply peaked that its internal structure, especially
the squeezed nature, is not easy to see.

Finally, Fig.~\ref{fig:type2-error} (left) shows the error statistics
for the $\omega$ and $\gamma$ estimates obtained using Strategy~3 for
10 models of type (\ref{eq:meas2}) with the same values for $\Omega$
and $\gamma$ as in Table~\ref{table:models}.  We compare two
experimental conditions: $\theta_I=\theta_M=0$, which corresponds to
maximum visibility of the oscillations and $\theta_I=\tfrac{\pi}{3}$,
$\theta_M=\tfrac{\pi}{4}$, for which the signal is more complex and
the visibility of the oscillations is reduced as shown in
Fig.~\ref{fig:type2}.  The estimation errors are very similar to those
for models of type 1.  For $\gamma$ they are effectively identical for
both experimental conditions; for $\Omega$ they are slightly larger in
case 2b, as might be expected as the visibility of the oscillations is
reduced in this case.

In both cases we also obtain excellent estimates of the noise level
$\sigma$ of the data as well as estimates for the parameters
$\alpha_1$ and $\alpha_2$.  As before, if the initial state prepared
or the precise measurement performed are unknown a priori, as may well
be the case for a system that is not yet well characterized, we can
use these parameters to derive estimates for $\theta_I$ and $\theta_M$:
\begin{subequations}
\begin{align}
\theta_I &= \frac{1}{2}[\arccos(\alpha_2-\alpha_1)+\arccos(\alpha_2+\alpha_1)]\\
\theta_M &= \frac{1}{2}[\arccos(\alpha_2-\alpha_1)-\arccos(\alpha_2+\alpha_1)]
\end{align}
\end{subequations}
Fig.~\ref{fig:type2-error} (right) shows the estimates derived for
the angles $\theta_I$ and $\theta_M$ for both experimental conditions.
The markers indicate the average of the estimate for all runs and all
model systems, the errorbars indicate the standard deviation of the
estimates. The estimates are not as accurate as those for the
system parameters, as one would expect as we have marginalized the
amplitudes $\alpha_1$ and $\alpha_2$ and thus $\theta_I$ and $\theta_M$.
However, they are still quite close to the actual values (black
dash-dot lines) with the exception of the $\theta_I$ estimate for
case (2a), which is slightly more biased and less accurate -- it should
be 0, coinciding with the measurement angle $\theta_M$.

\section{Conclusions}

We have investigated the ubiquitous problem of identifying crucial
parameters from experimental data for two-level systems subject to
decoherence.  Comparing different strategies based on the analysis of
Fourier spectra as well as Bayesian modelling and maximum likelihood
estimation, the latter approach was found to be vastly superior to
commonly used Fourier based strategies in terms of accuracy and
precision of the estimates obtained.

Strategies based on simple Fourier analysis are limited by the
accuracy with which the positions, heights and widths of the Fourier
peaks can be determined.  As the spectral resolution is limited by
signal length and sampling rate, the accuracy of Fourier-based
estimation schemes for short, decaying signals or sparse noisy data is
limited.  The Bayesian approach is not constrained in this way and
yields uncertainties for the system parameters as well as information
about the noise in the data.

An additional advantage of the Bayesian estimation is that it does not
require a priori knowledge of the initialization or measurement angles
$\theta_I$ and $\theta_M$.  Rather, the estimation procedure provides
values for the coefficients of the basis functions, which are related
to the parameters $\theta_I$ and $\theta_M$.

The results are widely applicable to many experimental settings from
the analysis for free-induction decay signals for spin systems, e.g.,
in NMR, MRI and ESR to Rabi spectrocopy fo atomic ensembles, trapped
ions, quantum dots or Josephson junction devices.

\acknowledgements We acknowledge funding from the Ser Cymru National
Research Network in Advanced Engineering and Materials.  SGS also
thanks the Royal Society for funding through a Leverhulme Senior
Fellowship grant and the UK Engineering and Physical Sciences Research
Council for recent funding.  FCL acknowledges funding from the Cardiff
University Research Leave Fellowship Scheme.

\end{document}